\documentclass[fleqn,usenatbib]{mnras}

\usepackage{newtxtext,newtxmath}

\usepackage[T1]{fontenc}
\usepackage{ae,aecompl}

\usepackage{graphicx}	
\usepackage{amsmath}	
\usepackage{amssymb}	
\usepackage{subcaption}
\usepackage{pdflscape}
\usepackage{multirow}
\usepackage{rotating}
\usepackage[flushleft]{threeparttable}
\usepackage{color}
\usepackage{soul}
\usepackage{lipsum}

\title[SCUBA-2 observations of starbursting protoclusters]{SCUBA-2 observations of candidate starbursting protoclusters selected by \textit{Planck} and \textit{Herschel}-SPIRE}

\author[T. Cheng et al.]{
T. Cheng,$^{1}$\thanks{E-mail: t.cheng15@imperial.ac.uk}
D.L. Clements,$^{1}$
J. Greenslade,$^{1}$
J. Cairns$^{1}$
P. Andreani,$^{2}$
M. Bremer,$^{3}$
\newauthor{
L. Conversi,$^{4}$
A. Cooray,$^{5}$
H. Dannerbauer,$^{6,7}$
G. De Zotti,$^{8}$
S. Eales,$^{9}$
}
\newauthor{
J. Gonz\'{a}lez-Nuevo,$^{10}$
E. Ibar,$^{11}$
L. Leeuw,$^{12}$
J. Ma,$^{5}$
M. J. Micha\l owski,$^{13}$
}
\newauthor{
H. Nayyeri,$^{5}$
D. A. Riechers,$^{14, 15}$
D. Scott,$^{16}$
P. Temi,$^{17}$
M. Vaccari,$^{18, 19}$
}
\newauthor{
I. Valtchanov,$^{20}$
E. van Kampen$^{2}$
and
L. Wang$^{21, 22}$
}
\\
$^{1}$Astrophysics Group, Imperial College London, Blackett Laboratory, Prince Consort Road, London SW7 2AZ, UK\\
$^{2}$European Southern Observatory, Karl-Schwarzschild-Stra${\ss}$e 2, 85748 Garching, Germany\\
$^{3}$HH Wills Physics Laboratory, University of Bristol, Tyndall Avenue, Bristol, BS8 1TL, UK\\
$^{4}$European Space Agency / ESAC, Camino Bajo del Castillo, 28692 Villanueva de la Ca\~nada (Madrid), Spain\\
$^{5}$Department of Physics and Astronomy, University of California, Irvine, CA 92697, USA\\
$^{6}$Instituto de Astrof\'{i}sica de Canarias, E-38205 La Laguna, Tenerife, Spain\\
$^{7}$Universidad de La Laguna Dpto. Astrof\'{i}sica, E-38206 La Laguna, Tenerife, Spain\\
$^{8}$INAF-Osservatorio astronomico di Padova, Vicolo dell'Osservatorio 5, I-35122 Padova, Italy\\
$^{9}$School of Physics and Astronomy, Cardiff University, The Parade, Cardiff CF24 3AA, UK\\
$^{10}$Departamento de F\'{i}sica, Universidad de Oviedo, C. Federico Garc\'{i}a Lorca 18, 33007 Oviedo, Spain\\
$^{11}$Instituto de F\'isica y Astronom\'ia, Universidad de Valpara\'iso, Avda. Gran Breta\~na 1111, Valpara\'iso, Chile\\
$^{12}$College of Graduate Studies, University of South Africa, PO Box 392, UNISA, 0003, South Africa\\
$^{13}$Astronomical Observatory Institute, Faculty of Physics, Adam Mickiewicz University, ul.~S{\l}oneczna 36, 60-286 Pozna{\'n}\\
$^{14}$Department of Astronomy, Cornell University, Space Sciences Building, Ithaca, NY 14853, USA\\
$^{15}$Max-Planck-Institut f\"ur Astronomie, K\"onigstuhl 17, D-69117 Heidelberg, Germany\\
$^{16}$Department of Physics and Astronomy, University of British Columbia,Vancouver, BC V6T1Z1, Canada\\
$^{17}$Astrophysics Branch, NASA Ames Research Center, Moffett Field, CA 94035, USA\\
$^{18}$Department of Physics and Astronomy, University of the Western Cape, Private Bag X17, Bellville 7535, Cape Town, South Africa\\
$^{19}$INAF - Istituto di Radioastronomia, via Gobetti 101, 40129 Bologna, Italy\\
$^{20}$Telespazio Vega UK for ESA, European Space Astronomy Centre, Operations Department, E-28691 Villanueva de la Ca\~nada, Spain\\
$^{21}$SRON Netherlands Institute for Space Research, Landleven 12, 9747 AD, Groningen, The Netherlands\\
$^{22}$Kapteyn Astronomical Institute, University of Groningen, Postbus 800, 9700 AV, Groningen, The Netherlands
}

\date{Accepted 2019 September 17. Received 2019 August 19; in original form 2019 April 25}

\date{Accepted 2019 September 17. Received 2019 August 19; in original form 2019 April 25}

\pubyear{2019}

\begin{document}
\label{firstpage}
\pagerange{\pageref{firstpage}--\pageref{lastpage}}
\maketitle

\begin{abstract}

We present SCUBA-2 850-$\mu$m observations of 13 candidate starbursting protoclusters selected using \emph{Planck} and \emph{Herschel} data. The cumulative number counts of the 850-$\mu$m sources in 9/13 of these candidate protoclusters show significant overdensities compared to the field, with the probability $<$10$^{-2}$ assuming the sources are randomly distributed in the sky. Using the 250-, 350-, 500- and 850-$\mu$m flux densities, we estimate the photometric redshifts of individual SCUBA-2 sources by fitting spectral energy distribution (SED) templates with an MCMC method. The photometric redshift distribution, peaking at $2<z<3$, is consistent with that of known $z>2$ protoclusters and the peak of the cosmic star-formation rate density (SFRD). We find that the 850-$\mu$m sources in our candidate protoclusters have infrared luminosities of $L_{\mathrm{IR}}\gtrsim$10$^{12}L_{\odot}$ and star-formation rates of SFR=(500-1,500)$M_{\odot}$yr$^{-1}$. By comparing with results in the literature considering only \emph{Herschel} photometry, we conclude that our 13 candidate protoclusters can be categorised into four groups: six of them being high-redshift starbursting protoclusters, one being a lower-redshift cluster/protocluster, three being protoclusters that contain lensed DSFG(s) or are rich in 850-$\mu$m sources, and three regions without significant \emph{Herschel} or SCUBA-2 source overdensities. The total SFRs of the candidate protoclusters are found to be comparable or higher than those of known protoclusters, suggesting our sample contains some of the most extreme protocluster population. We infer that cross-matching \emph{Planck} and \emph{Herschel} data is a robust method for selecting candidate protoclusters with overdensities of 850-$\mu$m sources.

\end{abstract}




\section{Introduction}

Protoclusters are defined as structures that are expected to collapse into galaxy clusters before the present epoch \citep[e.g.][]{2016A&ARv..24...14O}. Normally at $z>$2 , their hot gas halos may not yet be virialized. Without the virialized hot gas, protoclusters are difficult to find via traditional cluster-detection methods, such as X-rays or the Sunyaev-Zeldovich Effect (SZE). They also may not have a significant number of red sequence galaxies which would allow them to be identified on the optical colour-magnitude diagram (CMD), where red sequence galaxies at a fixed redshift cluster together due to similar stellar populations \citep{2007ApJ...671L..93B, 2008ApJ...684..905E, 2014ApJ...788...51N, 2014A&A...565A.120A}. Recent observations of high-$z$ (proto)clusters have found some red sequence galaxies \citep{2018arXiv181007330M}, but such cases are rare.

Most protoclusters are found through optical/near-infrared surveys, using overdensities of Lyman-$\alpha$ emitters (LAEs), Lyman-break galaxies (LBGs) or H-$\alpha$ emitters (HAEs) identified in the field \citep[e.g. The Hyper Suprime-Cam Subaru Strategic Program, HSC-SSP:][]{2018PASJ...70S...8A, 2018PASJ...70S...4A, 2010MNRAS.409.1155D} or around rare objects such as QSOs or radio galaxies \citep{2000A&A...361L..25P, 2004A&A...428..817K, 2008A&A...491...89V, 2011PASJ...63S.415T, 2012ApJ...757...15H, 2013MNRAS.432.2869H, 2015ApJ...808L..33C}. Thus QSOs and radio galaxies are often used as beacons when searching for protoclusters.

Dusty star-forming galaxies (DSFGs) are galaxies heavily obscured by dust and forming stars rapidly \citep{1997ApJ...490L...5S, 1998Natur.394..241H}. Submillimetre galaxies (SMGs) are a subsample of DSFGs selected in submillimetre surveys. The brightest and most luminous of these DSFGs can have luminosities exceeding $10^{13} \mathrm{L}_{\odot}$ \citep{2005ApJ...622..772C, 2015MNRAS.451.3419G, 2018A&A...619A.169R, 2018MNRAS.477.2042H}, which corresponds to star-formation rates (SFRs) of thousands of solar masses per year \citep{2014PhR...541...45C}, under standard calibrations \citep{2018Natur.558..260Z}. In galaxy formation models, DSFGs are thought to be the progenitors of elliptical, early-type galaxies residing in the cores of today's massive galaxy clusters \citep{2006ApJ...641L..17F, 2006ApJ...650...42L, 2008ApJ...689L.101F, 2010MNRAS.402.2113C, 2011ApJ...742...24L, 2013MNRAS.431..648W, 2013ApJ...768...21C, 2014ApJ...782...69L, 2014ApJ...782...68T, 2015ApJ...810...74A, 2017MNRAS.464.1380W}, so in principle we should see DSFGs in cluster progenitors, such as protoclusters.

Although it is statistically possible for line-of-sight overdensities of DSFGs to occur \citep{2017MNRAS.470.2253N}, there are existing observations of protoclusters hosting DSFGs \citep{2009ApJ...691..560C, 2009ApJ...694.1517D, 2014A&A...570A..55D, 2015ApJ...815L...8U, 2015ApJ...808L..33C} with SFRs as high as 3,000 $\mathrm{M}_{\odot}$yr$^{-1}$. Some of the DSFGs in protoclusters are also found to be formed of multiple sources by using higher resolution imagers, e.g. ALMA, VLT and HST \citep{2015ApJ...812...43B, 2016ApJ...828...56W, 2018ApJ...856..121G, 2018ApJ...856...72O, 2018arXiv180406581K, 2019ApJ...872..117G}. There are also cases where overdensities of DSFGs in the sky are line-of-sight projections of two protocluster structures at different redshifts \citep{2016A&A...585A..54F}. The redshifts of these protoclusters range from $z \simeq$2 to $z \simeq$5 \citep{2011Natur.470..233C, 2012Natur.486..233W, 2013ApJ...776...22H}, and their angular sizes vary from $<$1 arcminute to around 0.5 degree, probing regimes of size from cluster cores to large-scale structures such as filaments \citep{2004AJ....128.2073H, 2005ApJ...634L.125M, 2013ApJ...779..127C, 2016ApJ...828...56W, 2018ApJ...856...72O}.

According to some galaxy-formation models, during the formation of a galaxy cluster, member galaxies are expected to undergo a starbursting phase \citep{2004ApJ...600..580G, 2016ApJ...824...36C}, making them so-called ``starbursting galaxies". The timescale of this starbursting phase is short compared to the formation of a galaxy cluster and this phase might start and end at different times for different member galaxies; therefore, the probability that we observe a protocluster hosting a large number of DSFGs at the same time is very low \citep{2013ApJ...779..127C, 2016ApJ...824...36C}. The fact that a number of protoclusters containing DSFGs have been observed \citep{1998ApJ...492..428S, 2009ApJ...691..560C, 2014A&A...570A..55D, 2015ApJ...808L..33C} suggests there is an inconsistency between model predictions and observations.

Some models suggest that these starbursting galaxies might be mainly driven by mergers \citep{1985MNRAS.214...87J, 2008ApJS..175..356H, 2016MNRAS.462.2418S}, perhaps explaining why there are more DSFG-rich protoclusters than predicted. Alternatively, \cite{2016ApJ...824...36C} proposes that the starbursting phase in cluster galaxies can happen simultaneously based on observations of gas depletion times of known protoclusters, and hence, might explain why there are observations of DSFG-rich protoclusters.

Even though some optical surveys have been dedicated to protocluster searches, the total number of confirmed protoclusters to date is still limited to approximately 40 \citep{2016A&ARv..24...14O}. More protoclusters are needed, especially at high redshifts, in order to understand this population, and to resolve the inconsistency between observations and models. Searches for protoclusters in the submillimetre or far-infrared (FIR) bands are especially important because they directly look for DSFGs, and a galaxy's flux density can be as bright at $z\sim$8 as at $z=0$, due to the negative submm K-correction \citep[see][]{2014PhR...541...45C}.

\cite{2005MNRAS.358..869N} proposed a method that uses two FIR imaging instruments, one low and another high resolution, to identify candidate protoclusters. They show that the total flux density within the beam of a low-resolution instrument is in fact often the sum of a clump of several structures, which can be distinguished with a high-resolution instrument. Using this method with the \textit{Planck}\footnote{\textit{Planck} (\url{http://www.esa.int/Planck}) is a project of the European Space Agency (ESA) with instruments provided by two scientific consortia funded by ESA member states (in particular the lead countries France and Italy), with contributions from NASA (USA) and telescope reflectors provided by a collaboration between ESA and a scientific consortium led and funded by Denmark.} High Frequency Instrument \citep[HFI,][]{2010A&A...520A...9L} as the low resolution instrument and \textit{Herschel}\footnote{\textit{Herschel} \citep{2010A&A...518L...1P} is an ESA space observatory with science instruments provided by European-led Principal Investigator consortia and with important participation from NASA.} SPIRE \citep[Spectral and Photometric Imaging Receiver,][]{2010A&A...518L...3G} as the high resolution instrument, a number of candidate protoclusters have been found \citep{2013A&A...549A..31H, 2014MNRAS.439.1193C, 2016MNRAS.461.1719C, 2016A&A...596A.100P, 2018MNRAS.476.3336G} using maps from H-ATLAS \citep{2010PASP..122..499E} and HerMES \citep{2012MNRAS.424.1614O} surveys \citep[as well as targeted observations,][]{2016A&A...594A..27P}. This method is ideal for searching for high-redshift protoclusters, probing their formation periods and covering the epoch where the cosmic star-formation density peaks \citep{2006ApJ...651..142H, 2014MNRAS.439.1193C}.

Even though a number of candidate protoclusters have been identified using this method, their total cluster flux densities are found to be greater than in simulations by about a factor of 3 at 350 $\mu$m \citep{2015MNRAS.450.1320G, 2018MNRAS.476.3336G}. Such an inconsistency between observations and simulations in \cite{2015MNRAS.450.1320G} cannot simply be explained by starbursts with higher SFRs, since this will not give the correct total cluster stellar mass expected at $z$=0 in the simulations. In order to resolve this issue, observations need to be conducted on more protoclusters in order to feed into simulations.

Follow-up observations of candidate protoclusters selected using the above method have already been conducted. Observations using LABOCA \citep[LArge APEX BOlometer CAmera,][]{2009A&A...497..945S} and SCUBA-2 \citep{2013MNRAS.430.2513H} have found overdensities of submillimetres sources in a candidate protocluster which hosts a lensed DSFG at z=3.26 \citep{2016MNRAS.461.1719C}. Photometric redshifts of a few other candidate protoclusters using ancillary data suggest redshifts from 1 to 3 \citep{2014MNRAS.439.1193C}. In this paper we present follow-up observations of 13 candidate protoclusters selected using the above method. We study their number counts, far-infrared colours and photometric redshifts using observations from SCUBA-2 and \textit{Herschel}-SPIRE.

The content of this paper is as follows. In Section \ref{observations} we describe the \textit{Herschel}-SPIRE and SCUBA-2 observations of these 13 candidate protoclusters. In Section \ref{Results} we present results on number counts, FIR colours and photometric redshifts. We discuss and conclude in Sections \ref{Discussion} and \ref{Conclusions}, respectively. The standard concordance cosmology of H$_{0} = 70 $kms$^{-1}$Mpc$^{-1}$, $\Omega_{\mathrm{M}} = 0.3$, and $\Omega_{\Lambda} = 0.7$ is used throughout this paper.

\section{Observations and Data Reduction}\label{observations}

\subsection{Selection of candidate protoclusters}

The selection process for the candidate protoclusters, including those observed with SCUBA-2 described in this paper, is defined in \cite{2018MNRAS.476.3336G} (hereafter Gr18). Here we briefly summarise the selection process.

Using a protocluster survey technique, as introduced in \cite{2005MNRAS.358..869N}, sources from \textit{Planck} catalogues of compact sources (beam FWHM 5$\arcmin$) were selected. Subsequently these were examined in \textit{Herschel}-SPIRE maps (beam FWHM 18--36$\arcsec$) and catalogues from H-ATLAS and HerMES surveys to exclude local galaxies, Galactic cirrus or lensed objects. SPIRE has bands at 250-, 350-, and 500-$\mu$m, and is often used to select dusty and far-infrared-bright sources at $z>$2 \citep{2013Natur.496..329R, 2014ApJ...780...75D, 2016MNRAS.462.1989A} or lensed DSFGs \citep{2013ApJ...762...59W, 2016ApJ...823...17N}. 

After the above selection criteria, some of the remaining \textit{Planck} compact sources \citep{2013A&A...549A..31H} were targeted with other observations \citep{2014MNRAS.439.1193C, 2016MNRAS.461.1719C}. Gr18 further identified 27 \textit{Planck} compact sources as candidate high-redshift protoclusters where their \textit{Herschel} source overdensities were $>$3$\sigma$ in at least one SPIRE band.

Starting with the \textit{Planck} catalogues of compact sources, we looked for overdensities of \textit{Herschel}-SPIRE sources in the 250, 350, and/or 500 $\mu$m bands (\cite{2018MNRAS.476.3336G}) from H-ATLAS (\cite{2010PASP..122..499E}) and HerMES (\cite{2012MNRAS.424.1614O}) surveys. One of the aims of these surveys are to look at fields in the sky which are also covered by other multi-wavelength and extragalactic surveys. The total area of sky covered by H-ATLAS and HerMES are 570 $deg^{2}$ and 70 $deg^{2}$, respectively. The instrument we use is SPIRE, which has bands in 250, 350, and 500 $\mu$m, covering the far-infrared range and is often used to select dusty and far-infrared-bright sources at z$>$2 (\cite{2013Natur.496..329R}; \cite{2014ApJ...780...75D}; \cite{2016MNRAS.462.1989A}). Such dusty sources have rest-frame spectral energy distributions (SEDs) peaking at $\sim$100 $\mu$m due to re-radiated dust-obscured UV lights, and these dust-SED peaks are redshifted to SPIRE bands.

The SCUBA-2 observations were carried out for 13 fields of candidate protoclusters selected in Gr18 which are accessible by JCMT. Their properties (coordinates and overdensities) are listed in Table \ref{Table1}. The overdensity values are reported from Gr18, in which they counted the number of \textit{Herschel}/SPIRE sources with flux densities above 25.4 mJy at 350 or 500$\mu$m within the \textit{Planck} beam and compared with the expected number counts from \cite{2010A&A...518L...8C} and \cite{2016MNRAS.462.3146V}.

Note that five candidate protocluster fields were in fact not classified as candidate protoclusters in Gr18 because their \textit{Herschel}-SPIRE source overdensities are below 3$\sigma$. However, we still describe them as candidate protoclusters throughtout this paper, since Gr18 only applied their 3$\sigma$ overdensity cut to sources in \textit{Herschel}-SPIRE bands; protocluster galaxies that are brighter at other bands (particularly 850 $\mu$m), might be missed. Section \ref{selection_effect} discusses this additional selection effect.

The specific candidate protoclusters labelled Bootes1, EGS and Lockman were studied in \cite{2014MNRAS.439.1193C} and have redshift estimates of $2.27 \pm 0.24$, $0.76 \pm 0.10$, and $2.05 \pm 0.09$, respectively. The candidate protocluster G12 is also well studied \citep{2016MNRAS.461.1719C} and is believed to be at $z = 3.26$, consistent with the lensed DSFG in the centre of the field, which has a spectroscopic redshift \citep{2012ApJ...753..134F}.

\begin{table*}
\centering
\small
\caption{List of 13 candidate protoclusters observed with SCUBA-2 in this paper. Overdensity values are quoted from Gr18, in which \textit{Herschel} sources were selected with flux densities above 25.4 mJy at 350 or 500 $\mu$m within the \textit{Planck} beam, and compared with expected field number counts from \protect\cite{2010A&A...518L...8C} and \protect\cite{2016MNRAS.462.3146V}. Note that G12, NGP2, NGP3, NGP6 and NGP9 have overdensities below 3$\sigma$ in all SPIRE bands so were not selected as candidate protoclusters in Gr18 but we still refer to them as candidate protoclusters in this paper. The last column shows the category after comparing the overdensities of candidate protoclusters in Gr18 and in this paper, which will be discussed in Section \ref{selection_effect}.}
\label{Table1}
\begin{tabular}{ccccccc}
\hline Name    & \begin{tabular}[c]{@{}c@{}}RA\\ (J2000)\end{tabular} & \begin{tabular}[c]{@{}c@{}}Dec\\ (J2000)\end{tabular} & \begin{tabular}[c]{@{}c@{}}Overdensity at \\ 250 $\mu$m ($\sigma_{250}$)\end{tabular} & \begin{tabular}[c]{@{}c@{}}Overdensity at \\ 350 $\mu$m ($\sigma_{350}$)\end{tabular} & \begin{tabular}[c]{@{}c@{}}Overdensity at \\ 500 $\mu$m ($\sigma_{500}$)\end{tabular} & \begin{tabular}[c]{@{}c@{}}Category\\ (Section \ref{selection_effect})\end{tabular} \\ \hline
Bootes1 & 14:34:18.1                                           & +35:33:20.0                                           & 1.0                                                                                   & 4.5                                                                                   & 5.7                                                                                   & I                                                                                   \\
EGS     & 14:24:35.8                                           & +52:56:42.0                                           & 4.7                                                                                   & 5.8                                                                                   & 2.4                                                                                   & I                                                                                   \\
G12     & 11:46:33.6                                           & -00:11:15.0                                           & 0.8                                                                                   & 2.8                                                                                   & 2.4                                                                                   & III                                                                                 \\
Lockman & 10:33:26.9                                           & +59:10:09.1                                           & 5.4                                                                                   & 4.5                                                                                   & 4.7                                                                                   & II                                                                                  \\
NGP1    & 13:24:25.5                                           & +28:44:47.6                                           & 2.3                                                                                   & 2.0                                                                                   & 3.3                                                                                   & IV                                                                                  \\
NGP2    & 13:19:37.2                                           & +26:28:01.6                                           & 1.0                                                                                   & 2.2                                                                                   & 2.9                                                                                   & III                                                                                 \\
NGP3    & 13:31:42.9                                           & +23:46:16.9                                           & 1.5                                                                                   & 1.4                                                                                   & 1.6                                                                                   & III                                                                                 \\
NGP4    & 13:14:26.0                                           & +26:30:35.6                                           & 2.5                                                                                   & 3.3                                                                                   & 4.0                                                                                   & I                                                                                   \\
NGP5    & 13:40:41.4                                           & +32:37:17.1                                           & 2.1                                                                                   & 3.5                                                                                   & 3.3                                                                                   & I                                                                                   \\
NGP6    & 13:23:12.2                                           & +33:23:11.9                                           & 2.3                                                                                   & 0.8                                                                                   & 1.1                                                                                   & IV                                                                                  \\
NGP7    & 13:37:06.7                                           & +32:07:55.3                                           & 0.8                                                                                   & 2.5                                                                                   & 3.3                                                                                   & I                                                                                   \\
NGP8    & 13:29:26.2                                           & +28:13:25.4                                           & 1.7                                                                                   & 3.8                                                                                   & 4.0                                                                                   & I                                                                                   \\
NGP9    & 12:59:15.5                                           & +31:35:40.7                                           & 0.5                                                                                   & 1.7                                                                                   & 2.9                                                                                   & IV                                                                               \\  \hline
\end{tabular}
\end{table*}

\subsection{SCUBA-2 Observations}

\begin{figure}
\includegraphics[width=\columnwidth]{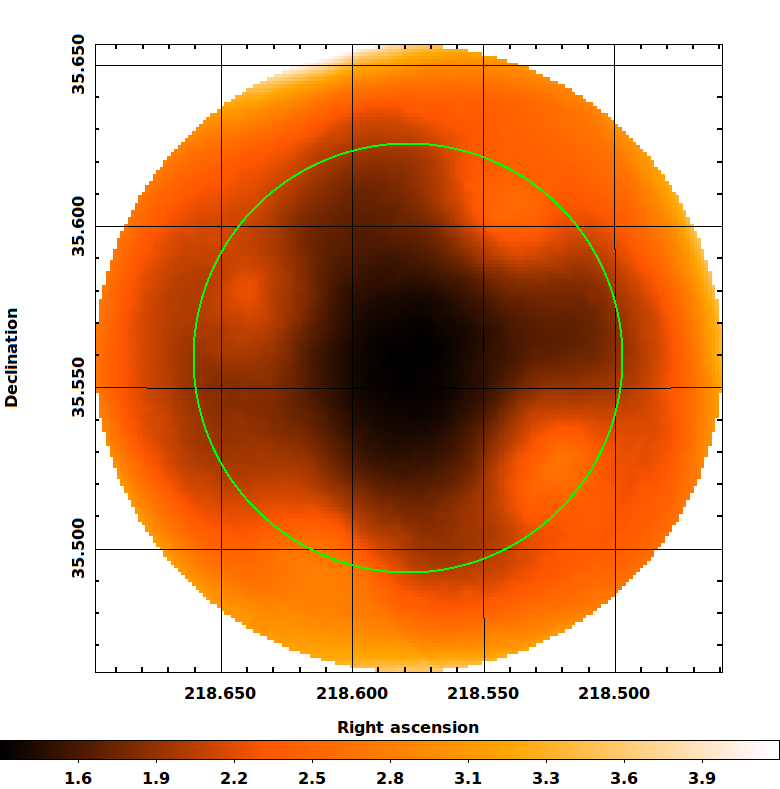}
\caption{The noise map of the candidate protocluster Bootes1. Colorbar units are in mJy. The green circle indicates the central region with 4 arcminute radius where the sensitivity is highest. North is up and east is left.}\label{variance}
\end{figure}

The JCMT/SCUBA-2 observations for these 13 candidate protoclusters took place between 8 and 12 April, 2013 (Project number M13AU12, PI D. Clements). The data were obtained at both 450 and 850 $\mu$m simultaneously, but the 450-$\mu$m maps did not reach the sensitivity needed to detect protocluster galaxies due to the weather conditions, so we only study the 850-$\mu$m maps here. Each candidate protocluster field was observed for approximately 2 hours, with several pointings using the CV Daisy mode \citep{2013MNRAS.430.2513H}, giving rms noise levels of approximately 2.25 mJy/beam. Weather conditions were good throughout these observations (precipitable water vapour between 0.83 mm and 2.58 mm) and standard calibrations were conducted. Fig.\ref{variance} shows an example of the noise distribution in one of our candidate protoclusters. The green circle shows the central region with 4 arcminute radius, indicating the most sensitive region.

The data were reduced using the SCUBA-2 pipeline \verb|SMURF| package \citep{2013MNRAS.430.2545C}. The raw SCUBA-2 data are composed of 30-second chunks of observations, for each of the four sub-arrays at each wavelength. They are in the STARLINK NDF format (.sdf). The first step is to use the \verb|MAKEMAP| command to create a map combining all sub-arrays and observing chunks for each scan. Under the \verb|MAKEMAP| command, the \verb|ITERATE| method is used, which fits models for the noise and the instrumental behaviour. Next, the \verb|PICARD| recipe \verb|MOSAIC_JCMT_IMAGES| is used, which co-adds separate scans into a single map and removes contaminant signals such as cosmic rays.

Since we expect these 850-$\mu$m sources to be point sources and unresolved due to their high redshifts, we used \verb|PICARD|'s \verb|SCUBA2_MATCHED_FILTER| recipe. This recipe first subtracts the background by convolving (smoothing) the maps and the PSF with a 30" FWHM Gaussian kernel. Then the signal maps are convolved with the PSF to produce the matched-filtered signal map. The noise maps are also convolved with the PSF to produce the variance map. This process essentially performs the maximum likelihood fit of the PSF on every pixel of the map, and is beneficial at finding sources with angular scales of the telescope beam. This process gives an effective beam FWHM of 14.6 arcseconds \citep{2013MNRAS.430.2534D} surrounded by a shallow negative ring.

The last step in the reduction is to crop the maps to a diameter of 700 arcsec (11.7 arcmin) where they are most sensitive using \verb|PICARD|'s \verb|CROP_SCUBA2_IMAGES| recipe and produce signal-to-noise maps using the \verb|MAKESNR| command. The standard flux conversion factor (FCF) of 537 Jy/pW at 850 $\mu$m was used for the calibration from detected power units to flux density \citep{2013MNRAS.430.2534D}. According to JCMT/SCUBA-2 website\footnote{\url{https://www.eaobservatory.org/jcmt/instrumentation/continuum/scuba-2/calibration/}}, the calibration error is approximately 5 percent. We include the 5$\%$ calibration error and add it in quadrature to the uncertainties of the flux densities of the 850 $\mu$m sources.

\subsection{Source Detection}

In order to build our SCUBA-2 850-$\mu$m source catalogue, we find peaks in the S/N maps, which were obtained by dividing the beam-convolved map with the noise map. Nearby, connected pixels having S/N$>$3.5 are regarded as part of the same source and the highest S/N pixel is used to record the flux density and noise of the source.

A correction to the flux densities and noise is needed, due to flux boosting. We apply the correction from equation 5 of \cite{2017MNRAS.465.1789G}, in which the flux density and noise correction is a power-law function of the observed signal-to-noise ratio ($SNR$):

\begin{equation}
\frac{S_{obs}}{S_{true}} = 1 + 0.2 \left( \frac{SNR}{5} \right)^{-2.3},
\end{equation}

where $S_{obs}$ is the observed flux density and noise; $S_{true}$ is the de-boosted flux density and noise.

We study the reliability of the sources by inverting the flux density maps, which are background subtracted, and using the same extraction method as above. By doing this, negative sources due to negative noise peaks will be found. We assume there is a similar number of spurious sources in the original maps due to positive noise peaks. By calculating the ratio between the detected number of positive sources and the expected number of spurious sources, we find a reliability of 80 percent at 3.5$\sigma$, consistent with similar SCUBA-2 studies of candidate protoclusters, such as in \cite{2017MNRAS.468.4006M}.

Since the beam deviates from a Gaussian, the number of positive and negative noise spikes may be different, which may cause incorrect reliability estimates using the above method. Hence we also produce a series of jackknife maps, where the data are separated into two halves alternately. The data of one half is then inverted and co-add with the other half, producing what is essentially the maps with pure noise. By doing source extraction on these jackknife maps, we found the reliability of all 13 candidate protocluster fields being significantly better than 80 percent at 3.5$\sigma$.

Using a higher S/N detection threshold than 3.5$\sigma$ would improve the reliability of the sources while decreasing the total number of sources. According to our reliability studies there are chances of spurious sources in our catalogue. For example, those with photometric redshifts of $z>$6 (see Table \ref{Table2}; discussed in Section \ref{photo_z}) are likely to be spurious due to their bad photometry. To test how much change different S/N values will incur, we impose different S/N cuts up to 5 $\sigma$ and perform the cumulative number count analysis (discussed in Section \ref{number_counts}). We find no significant change in the conclusions we make in this paper.

The completeness of the sources is estimated by inserting fake sources from 2 to 20 mJy to the flux density maps of each candidate protocluster field, and then calculating the fration of these sources recovered using the same source detection algorithm. The shape of the fake sources is a circular 2D Gaussian, with the standard deviation being half of the SCUBA-2 beam FWHM at 850 $\mu$m, i.e. 14.6 arcsec. We found that although the completeness levels vary between different fields, all but three candidate protocluster fields show a completeness level above 50 percent at 8 mJy, after flux-deboosting. If we restrict the fake sources to be within the 4 arcmin radius, all candidate protocluster fields have a completeness level above 50 percent at $\geq$8 mJy. In the three fields (EGS, NGP3, NGP8) where completeness falls below 50 percent at 8 mJy, only 2 sources lie outside the 4 arcmin radius region, and only in NFP3 field. We note these 2 sources in the source catalogue in Table \ref{Table2} since they have a lower completeness level. Overall we conclude our SCUBA-2 sources have a reasonable completeness level above 8 mJy and calculate the probability of the observed number of sources above this flux density in the studies of number counts (Section \ref{number_counts}).

Maps showing the detected sources in all 13 candidate protocluster fields are shown in Fig \ref{Fig2}. Catalogues showing the de-boosted flux densities of the detected SCUBA-2 850-$\mu$m sources for all 13 candidate protocluster fields are given in Table \ref{Table2}.

\section{Results}\label{Results}

\subsection{Cumulative Number Counts of SCUBA-2 Sources}\label{number_counts}

In order to study the existence of overdensities of 850-$\mu$m sources, cumulative number counts are estimated for these 13 fields, based on their de-boosted flux densities. 

\begin{landscape}

\begin{figure}
\includegraphics[width=\columnwidth]{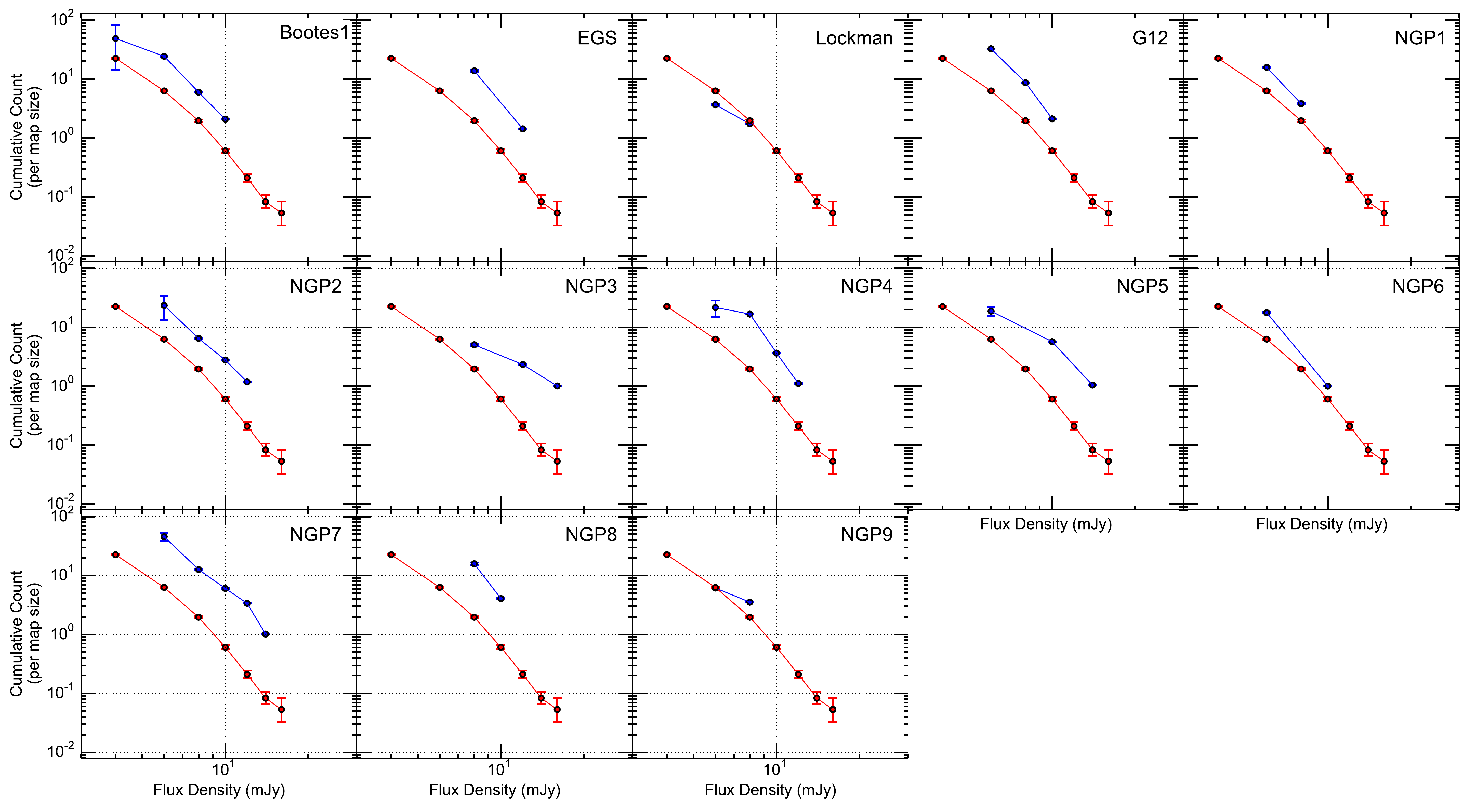}
\caption{Cumulative number counts of the SCUBA-2 sources in our candidate protocluster fields (blue) and in random fields (red) from \protect\cite{2017MNRAS.465.1789G} (Ge17). Note that the flux densities of the two known strongly lensed objects in G12 and NGP1 are de-magnified and considered. The errorbars of the SCUBA-2 sources in our candidate protocluster fields (blue) are the completeness error, and the errorbars of random fields (red) from Ge17 are Poissonian. The number of sources are scaled to the size of each candidate protocluster field, which is approximately 0.03 deg$^{2}$.}\label{Fig3}
\end{figure}

\end{landscape}

Fig.\ref{Fig3} shows the cumulative number counts. We sort the flux densities of the sources from bright to faint and count the cumulative number of sources at each 2-mJy step, and plot as blue points. The results from \cite{2017MNRAS.465.1789G} (hereafter Ge17) are shown in red, which is from the SCUBA-2 Cosmology Legacy Survey (S2CLS) and is composed of approximately 5 deg$^{2}$ of blank field.

These cumulative counts need to be corrected for the changing sensitivity across the maps. The effective area to detect each source is smaller than the entire map due to the attenuated sensitivity near the edge. We have corrected for this overestimated area by dividing the number of sources by the effective area (rather than the entire map area) corresponding to different sensitivities. Hence the number of sources in Fig \ref{Fig3} and Table \ref{Table_ncounts} is greater than that listed in Table \ref{Table2}.

\begin{table*}
\centering
\caption{Cumulative number counts of our 13 candidate starbursting protoclusters and in the field from the S2CLS survey \protect\citep{2017MNRAS.465.1789G}. Cumulative number counts represent the number of sources with flux densities $>$4 mJy, $>$6 mJy, $>$8 mJy, $>$10 mJy and $>$12 mJy. The uncertainties in the candidate protocluster fields represent the completeness error and the uncertainties in the S2CLS survey represent Poissonian errors. The number of sources is corrected for the variable sensitivity and is compared with field surveys. Flux densities of the two lensed sources in G12 and NGP1 are de-magnified based on the magnification factor (see text) and counted. The last row shows the expected number counts scaled to the area of each map, approximately 0.03 deg$^{2}$, in order to compare with the actual number of sources in the maps, as discussed in the text. The $P_{random}$ column shows the probability of the observed number of sources compared with the expected number in \protect\cite{2017MNRAS.465.1789G} at 8 mJy, assuming the sources are randomly distributed. $N_{overdensity}$ and $P_{overdensity}$ columns are the number of overdense random regions and the overdense level (fraction of regions showing an overdensity, see text), after putting 10,000 random regions in the S2CLS/COSMOS field to study the potential cosmic variance effect.}
\label{Table_ncounts}
\begin{tabular}{ccccccccc}
\hline Name                                                       & $>$4 mJy      & $>$6 mJy              & $>$8 mJy               & $>$10 mJy     & $>$12 mJy     & \begin{tabular}[c]{@{}c@{}}$P_{random}$\\ (at 8 mJy)\end{tabular} & $N_{overdensity}$ & $P_{overdensity}$ \\ \hline
Bootes1                                                    & 49$\pm$35     & 24$\pm$0.4            & 6.0$\pm$0.07           & 2.1$\pm$0.02  & N/A           & 0.02                                                              & 920               & 0.092             \\
EGS1                                                       & 14$\pm$0.6    & 14$\pm$0.6            & 14$\pm$0.6             & 1.4$\pm$0.02  & 1.4$\pm$0.02  & $2.44 \times 10^{-8}$                                             & 0                 & $< 10^{-4}$       \\
G12                                                        & 33$\pm$0.3    & 33$\pm$0.3            & 8.7$\pm$0.09           & 2.1$\pm$0.02  & N/A           & $9.97 \times 10^{-4}$                                             & 318               & 0.0318            \\
Lockman                                                    & 3.7$\pm$0.08  & 3.7$\pm$0.08          & 1.7$\pm$0.02           & N/A           & N/A           & 0.86                                                              & 4422              & 0.4422            \\
NGP1                                                       & 16$\pm$0.2    & 16$\pm$0.2            & 3.8$\pm$0.04           & N/A           & N/A           & 0.32                                                              & 1825              & 0.1825            \\
NGP2                                                       & 24$\pm$10     & 24$\pm$10             & 6.5$\pm$0.1            & 2.8$\pm$0.03  & 1.2$\pm$0.01  & 0.02                                                              & 616               & 0.0616            \\
NGP3                                                       & 5.1$\pm$0.2   & 5.1$\pm$0.2           & 5.1$\pm$0.2            & 2.3$\pm$0.03  & 2.3$\pm$0.03  & 0.05                                                              & 920               & 0.092             \\
NGP4                                                       & 22$\pm$7      & 22$\pm$7              & 17$\pm$0.3             & 3.7$\pm$0.04  & 1.1$\pm$0.01  & $4.46 \times 10^{-11}$                                            & 0                 & $< 10^{-4}$       \\
NGP5                                                       & 19$\pm$3      & 19$\pm$3              & 5.7$\pm$0.06           & 5.7$\pm$0.06  & 1.0$\pm$0.01  & 0.05                                                              & 920               & 0.092             \\
NGP6                                                       & 18$\pm$0.3    & 18$\pm$0.3            & 1.0$\pm$0.01           & 1.0$\pm$0.01  & N/A           & 0.86                                                              & 7233              & 0.7233            \\
NGP7                                                       & 46$\pm$6      & 46$\pm$6              & 13$\pm$0.2             & 6.1$\pm$0.08  & 3.4$\pm$0.04  & $1.75 \times 10^{-7}$                                             & 17                & 0.0017            \\
NGP8                                                       & 16$\pm$0.9    & 16$\pm$0.9            & 16$\pm$0.9             & 4.1$\pm$0.06  & N/A           & $3.88 \times 10^{-10}$                                            & 0                 & $< 10^{-4}$      \\
NGP9                                                       & 6.1$\pm$0.1   & 6.1$\pm$0.1           & 3.5$\pm$0.03           & N/A           & N/A           & 0.32                                                              & 1825              & 0.1825            \\ \hline
\begin{tabular}[c]{@{}c@{}}S2CLS\\ (expected)\end{tabular} & 22.6$\pm$0.34 & $6.3^{+0.16}_{-0.15}$ & $1.97^{+0.09}_{-0.08}$ & 0.61$\pm$0.05 & 0.21$\pm$0.03 &                                                                   &                   &               \\   \hline
\end{tabular}
\end{table*}

The cumulative number counts of the 13 candidate protoclusters and of the field in S2CLS from Ge17 are listed in Table \ref{Table_ncounts}. We now describe the number count behaviour of each candidate protocluster field and calculate the probability\footnote{This is the \textit{upper tail} of the probability density function following a Poisson distribution, calculated using R function \texttt{ppois(observed-1, lambda=expected, lower=FALSE)}.} of the observed number of sources above 8 mJy, assuming that the sources are randomly distributed. All the observed number counts are quoted after the variable sensitivity is corrected for in the effective size of each individual map. In the last row of Table \ref{Table_ncounts} we list the expected number counts from Ge17 scaled to the size of each candidate protocluster field, which is approximately 0.03 deg$^{2}$.

\textbf{Bootes1}: The cumulative number counts show overdensities ($>5 \sigma$) of 850 $\mu$m sources at 6.0, 8.0 and 10.0 mJy, compared to Ge17. We observe 6 sources with flux densities above 8.0 mJy in the map. Compared to the expected number from Ge17, and assuming that the sources are randomly distributed, the probability of observing this number of sources in a random field ($P$($\geq$ 6)) is 0.02, following a Poisson distribution. 

\textbf{EGS}: An overdensity of 850-$\mu$m sources is seen from 8.0 to 12.0 mJy in the cumulative number counts. We observe 14 sources with flux densities $>$8.0 mJy in the map. The probability of observing this number of sources in a random field ($P$($\geq$ 14)) is $2.44 \times 10^{-8}$.

\textbf{Lockman}: There is no overdensity in the observed flux density range. There are 1.7 sources with flux densities above 8.0 mJy in the map. The probability of observing this number of sources in a random field ($P$($\geq$ 1.7)) is 0.86.

\textbf{G12}: Overdensities in cumulative number counts are seen from 6.0 to 10.0 mJy. We observe 8.7 sources with flux densities above 8.0 mJy in the map. The probability of observing this number of sources in a random field ($P$($\geq$ 8.7)) is $9.97 \times 10^{-4}$. In the central region of G12, there is a strongly lensed system, which is at $z$=3.26 \citep{2012ApJ...753..134F, 2013A&A...549A..31H}, and has an observed SCUBA-2 flux density of 79.8$\pm$4.2 mJy. The 850 $\mu$m flux density of this source is de-magnified according to the magnification factor $\mu$(880$\mu$m) = 7.6 $\pm$ 1.5 from \cite{2012ApJ...753..134F} and considered in the cumulative number counts. This candidate protocluster was already studied in \cite{2016MNRAS.461.1719C}, and is believed to have a similar redshift to the spectroscopically confirmed lensed DSFG at $z$=3.26. Note that the number of observed FIR sources presented here is higher than that in \cite{2016MNRAS.461.1719C} because the quoted number of sources has been corrected for changing sensitivity.

\textbf{NGP1}: There is a bright SCUBA-2 source in the central region of this candidate protocluster with a flux density of 42.8$\pm$2.5 mJy, which is a z=1.676 lensed source \citep{2013ApJ...779...25B, 2014ApJ...797..138C, 2016ApJ...829...21T}. The flux density of this source is de-magnified according to the magnification factor of $\mu$(dust) = 4.9 $\pm$ 1.8 from \cite{2016ApJ...829...21T} and considered in the cumulative number counts. Only slight overdensities can be seen at 6.0 and 8.0mJy in the cumulative number counts. We observe 3.8 sources with flux densities above 8.0 mJy in the map. The probability of observing this number of sources in a random field ($P$($\geq$ 3.8)) is 0.32.

\textbf{NGP2}: Overdensities are seen from 6.0 to 12.0mJy in the cumulative number counts. We observe 6.5 sources with flux densities above 8.0 mJy in the map. The probability of observing this number of sources in a random field ($P$($\geq$ 6.5)) is 0.02.

\textbf{NGP3}: Overdensities are seen at 8.0, 12.0 and 16.0 mJy for cumulative number counts. We observe 5.1 sources with flux densities above 8.0 mJy in the map. The probability of observing this number of sources in a random field ($P$($\geq$ 5.1)) is 0.05.

\textbf{NGP4}: Overdensities are seen from 6.0 to 12.0mJy in the cumulative number counts. We observe 17 sources with flux densities above 8.0 mJy in the map. The probability of observing this number of sources in a random field ($P$($\geq$ 17)) is $4.46 \times 10^{-11}$.

\textbf{NGP5}: Cumulative number counts show overdensities from 6.0 to 14.0 mJy. We observe 5.7 sources with flux densities above 8.0 mJy in the map. The probability of observing this number of sources in the field ($P$($\geq$ 5.7)) is 0.05.

\textbf{NGP6}: No overdensity is seen at 10.0 mJy and slight overdensities can be seen at 6.0 mJy, in the cumulative number counts.  We observe one source with a flux density above 8.0 mJy in the map. The probability of observing this number of sources in a random field ($P$($\geq$ 1)) is 0.86. We do not include the flux density bin of 4.0 mJy since the effective area correction becomes unreliable at fainter flux densities, and the number counts need further investigation.

Among the SCUBA-2 detected sources, NGP6.02 is a Herschel-SPIRE dropout, which has no \textit{Herschel} counterpart and is believed to be either a $z>6$ SMG or a cool $z=4$ DSFG (Greenslade et al., submitted). Whether this source is associated with any protocluster or its line-of-sight overlaps needs further investigation.

\textbf{NGP7}: The cumulative number counts show overdensities from 6.0 to 14.0mJy. We observe 13 sources with flux densities above 8.0 mJy in the map. The probability of observing this number of sources in a random field ($P$($\geq$ 13)) is $1.75 \times 10^{-7}$.

\textbf{NGP8}: Cumulative number counts show overdensities at 8.0 and 10.0 mJy. We observe 16 sources with flux densities above 8.0 mJy in the map. The probability of observing this number of sources in a random field ($P$($\geq$ 16)) is $3.88 \times 10^{-10}$.

\textbf{NGP9}: A slight overdensity is seen in at 8.0 mJy, in cumulative number counts. We observe 3.5 sources with flux densities above 8.0 mJy in the map. The probability of observing this number of sources in a random field ($P$($\geq$ 3.5)) is 0.32.

We have estimated the probability of obtaining the observed number of sources in our candidate protoclusters compared to the field assuming the sources are randomly distrubuted. However, in reality they may not be randomly distributed and may be clustering due to, for example, cosmic variance. We therefore perform a test where 10,000 regions with the same size as the protocluster candidate maps (i.e. $\sim$0.03 $deg^{2}$) are randomly placed in the S2CLS/COSMOS field. The same source extraction method is used and we count the number of detections in each random region. We then count the regions that are overdense, i.e. containing more than the observed number of sources in each candidate protocluster fields (e.g. 6 sources $>$8 mJy in Bootes1). We list the number of these overdense regions ($N_{overdensity}$) and the overdensity levels, i.e. the fraction of regions more overdense than the protocluster candidates ($P_{overdensity} = N_{overdensity}$/10,000), in Table \ref{Table_ncounts}.

We found that for the candidate protocluster fields where there are overdensities of 850 $\mu$m sources, i.e. Bootes1, EGS, G12, NGP2, NGP3, NGP4, NGP5, NGP7 and NGP8, the overdensity level is  $<$ 0.1. Hence even under the potential clustering effect due to cosmic variance, these fields still have significant overdensities of 850 $\mu$m sources. The result from \cite{2016MNRAS.461.1719C}, which used submm sources in G12, found the overdensity level of 2.5$\times 10^{-2}$ within the two arcmin radius, consistent with our result of 3.18$\times 10^{-2}$. 

The uncertainties in flux boosting correction and completeness correction may also affect the significance of number counts. Nonetheless, it is found that the flux boosting correction is consistent among different methods \citep{2017MNRAS.465.1789G}, and the uncertainties are generally within 1 mJy, especially at brighter flux density bins ($>$5 mJy). The completeness uncertainties are below 6 percent among all candidate protocluster fields, so should not have significant effects on the result of number counts and the overall conclusions made in this paper.

\subsection{SCUBA2-\textit{Herschel} Colours} \label{colours}

We estimate the 250-, 350-, and 500-$\mu$m flux densities of our SCUBA-2 sources in all 13 candidate protoclusters using photometry data from \textit{Herschel} in the H-ATLAS (Data Release 1 for G12 and Data Release 2 for NGP fields) and HerMES (Data Release 4\footnote{The HerMES source catalogues can be downloaded from HeDaM (\url{http://hedam.lam.fr}).}, except for Bootes1, where we use Data Release 2, due to missing 500-$\mu$m photometry information in DR4) surveys \citep{2012MNRAS.419..377S, 2010MNRAS.409...48R, 2014MNRAS.444.2870W}.

Firstly we match the positions of our SCUBA-2 sources with those in the H-ATLAS or HerMES catalogues, using an 18 arcsec search radius, which matches the SPIRE 250-$\mu$m beam from which the \textit{Herschel}-SPIRE catalogues are derived. We then add a random offset to the original SCUBA-2 source positions and then the same matching algorithm is conducted. We find that the search radius with minimum number of spurious matches is 9''. There are 10 SCUBA-2 sources which have matches in the \textit{Herschel} catalogues with separation between 9'' and 18''. We mark these 10 sources (as $cat*$) in the source catalogue in Table \ref{Table2}.

If no match is found for a SCUBA-2 source beyond 18'', we use the flux densities and noise from the maps at 250, 350 and 500 $\mu$m at the positions of the SCUBA-2 sources.

Using the above method we are able to constrain the \textit{Herschel}-SPIRE flux densities of each SCUBA-2 source in the candidate protocluster fields and extract the 4-band (250, 350, 500, and 850 $\mu$m) photometry. The 250-, 350-, 500- and 850-$\mu$m flux densities of all SCUBA-2 sources in all 13 candidate protocluster fields are listed in Table \ref{Table2}.

\begin{landscape}

\begin{figure}
\includegraphics[width=\columnwidth]{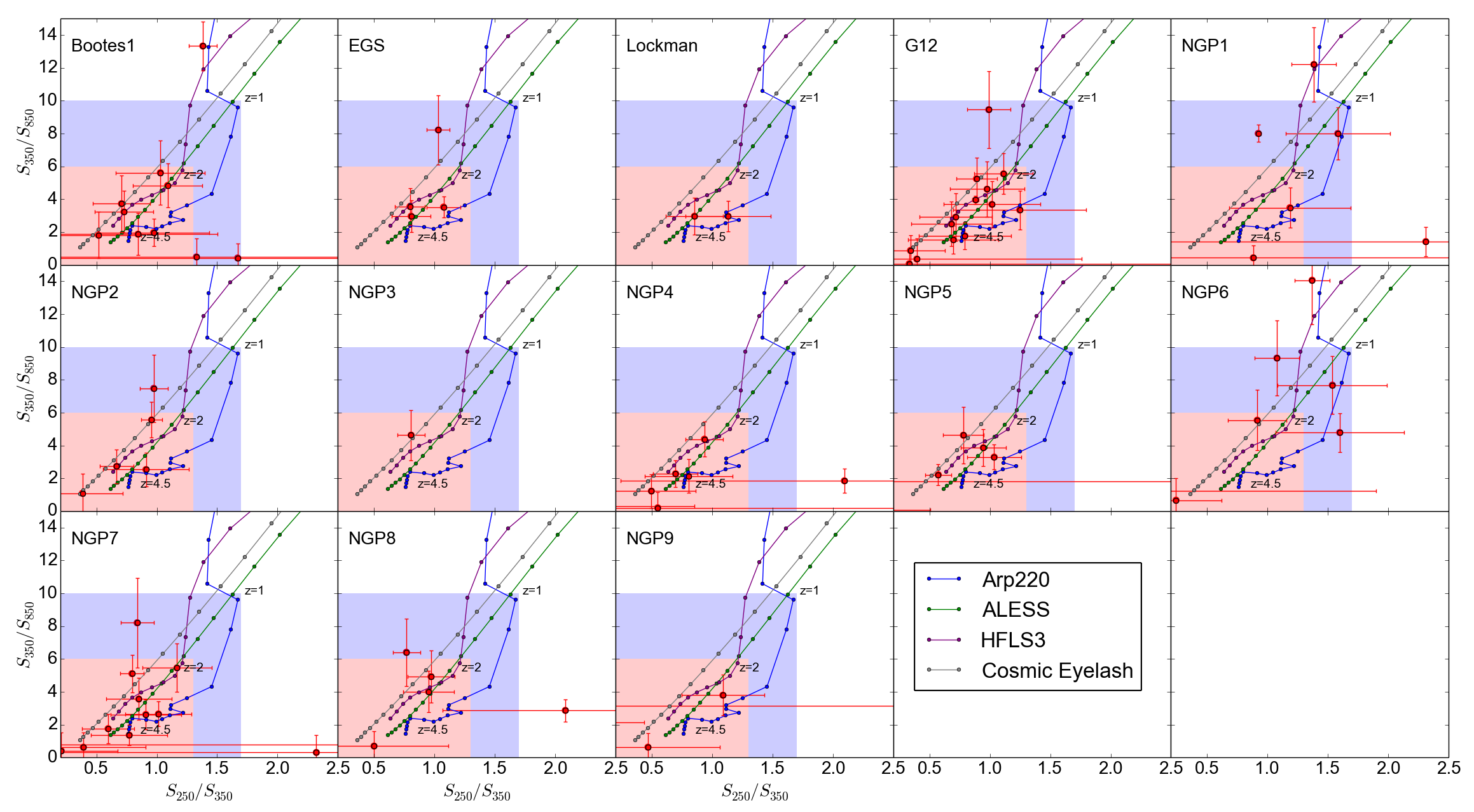}
\caption{Submillimetre colour-colour plots, specifically 250$\mu$m/350$\mu$m versus 350$\mu$m/850$\mu$m plots of 13 candidate protocluster fields observed with SCUBA-2 in this paper. Red points and error bars represent the 850-$\mu$m sources in the candidate protocluster fields. Errors are propagated from the 250-, 350-, and 850-$\mu$m flux densities of individual sources. Blue, green, purple and grey curves are the redshift tracks on this colour-colour plot from $z$=0.0 to 4.5 with 0.25 steps, using templates of Arp220, ALESS, HFLS3 and Cosmic Eyelash, respectively. Based on the redshift tracks of the template SEDs, we identify colours where sources are potentially at $z \geq 2$ as the red region, and colours where sources are potentially at $1 < z < 2$ as the blue region.}\label{Fig1}
\end{figure}

\end{landscape}

Once the 850-$\mu$m and SPIRE flux densities are obtained, we can derive the colours, a model-independent approach to identify high-redshift DSFGs. Figure \ref{Fig1} shows the 250$\mu$m/350$\mu$m versus 350$\mu$m/850$\mu$m colour-colour plots of the SCUBA-2 sources from the 13 candidate protocluster fields. We also plot the colours derived from template SEDs of the local ULIRG Arp220 \citep{2007ApJ...660..167D, 2011ApJ...743...94R}, average SMGs from the ALMA-LABOCA ECDFS Submm Survey (ALESS) \citep{2015ApJ...806..110D}, the high-z source HFLS3 \citep{2013Natur.496..329R} and the Cosmic Eyelash \citep{2010Natur.464..733S}, by redshifting these SEDs from $z$=0 to $z$=4.5. These template SEDs are representative of starbursting galaxies or SMGs at various redshifts, and Table \ref{table_template} summarises their properties.

\begin{table*}
\centering
\caption{Properties of the model SEDs used in this paper.}
     \begin{threeparttable}
\begin{tabular}{llll}
\hline Name of model SED & dust temperature  & redshift    & nature of source    \\\hline
Arp220            & 66 K              & $\sim$0.018 \tnote{a} & local ULIRG         \\
HFLS3             & $56^{+9}_{-12}$ K & 6.34        & starbursting galaxy \\
ALESS             & $\sim$40 K        & 1.33-6.12   & SMGs                \\
Cosmic Eyelash    & 30-60 K           & 2.3         & SMG              \\  \hline
\end{tabular}
        \begin{tablenotes}
            \item[a] \protect\cite{Vaucouleurs_book}.
        \end{tablenotes}
     \end{threeparttable}
\label{table_template}
\end{table*}

Based on the redshift track of the template SEDs, we identify the potential $z \geq 2$ sources to have colours $S_{250}/S_{350} \leq 1.3$ and $S_{350}/S_{850} \leq 6.0$, and shown as the red region in Fig.\ref{Fig1}. Similarly we identify potential $1 < z < 2$ sources to have colours as shown in the blue region in Fig.\ref{Fig1}.

Among the 13 candidate protocluster fields, Bootes1, G12 and NGP7 all show a number of sources having colours lying within the $z \geq$2 region. The EGS, Lockman, NGP2, NGP3, NGP4, NGP5, NGP8, NGP9 fields have fewer sources, but their colours still suggest $z \geq$2. Sources in NGP1 and NGP6, on the other hand, suggest redshifts between 1 and 2.

\subsection{SED fitting and photometric redshifts}\label{photo_z}

We use the SPIRE 250-, 350-, 500- and SCUBA-2 850-$\mu$m flux densities derived in Section \ref{colours} to estimate the photometric redshifts of the SCUBA-2 sources in the protocluster fields. We use the same model SEDs as in Table \ref{table_template} to perform the fitting, as these templates are representative of DSFGs with a variety of dust temperatures. Due to the lack of information on the dust temperature of our sources, we expect to obtain different redshift estimates when using different template SEDs. Using a warmer dust-temperature template SED (e.g. Arp220), the resulting redshift is higher, while using a colder template (e.g. ALESS), the resulting redshift is lower. This is referred to as the temperature-redshift degeneracy \citep{1999MNRAS.304..669B, 1999ASPC..191..255B}, which prevents accurate redshift estimates if observed flux densities only rely on the (redshifted) thermal dust (modified blackbody) SED. Nonetheless, for redder DSFGs (which describes the majority of our SCUBA-2 sources), this effect is minor and only a slight scatter is seen if different template SEDs are used \citep{2016ApJ...832...78I}. \cite{2016ApJ...832...78I} also suggest that even with a limited number of templates, the estimated redshifts of the sources can still be accurate.

We perform $\chi^{2}$-minimisation using \verb|emcee| \citep{2013PASP..125..306F}, which is a \verb|Python| implementation of the affine-invariant ensemble sampler for a Markov chain Monte Carlo (MCMC) procedure \citep{2010CAMCS...5...65G}.  We fit two parameters, redshift \textit{z} and the normalization factor \textit{a}. The normalization factor \textit{a} is the factor by which template flux density values are multiplied and is associated with the luminosity of a source. We constrain the range of parameters to be $0<z<15.0$ and $10^{-2}<a<10^{2}$, and place flat priors on them. The MCMC chains are composed of 100 walkers, each having 5,000 steps, and a 1,000-step burn-in time is allowed.

\begin{figure*}
    \begin{subfigure}[b]{1.0\textwidth}
        \includegraphics[width=\columnwidth]{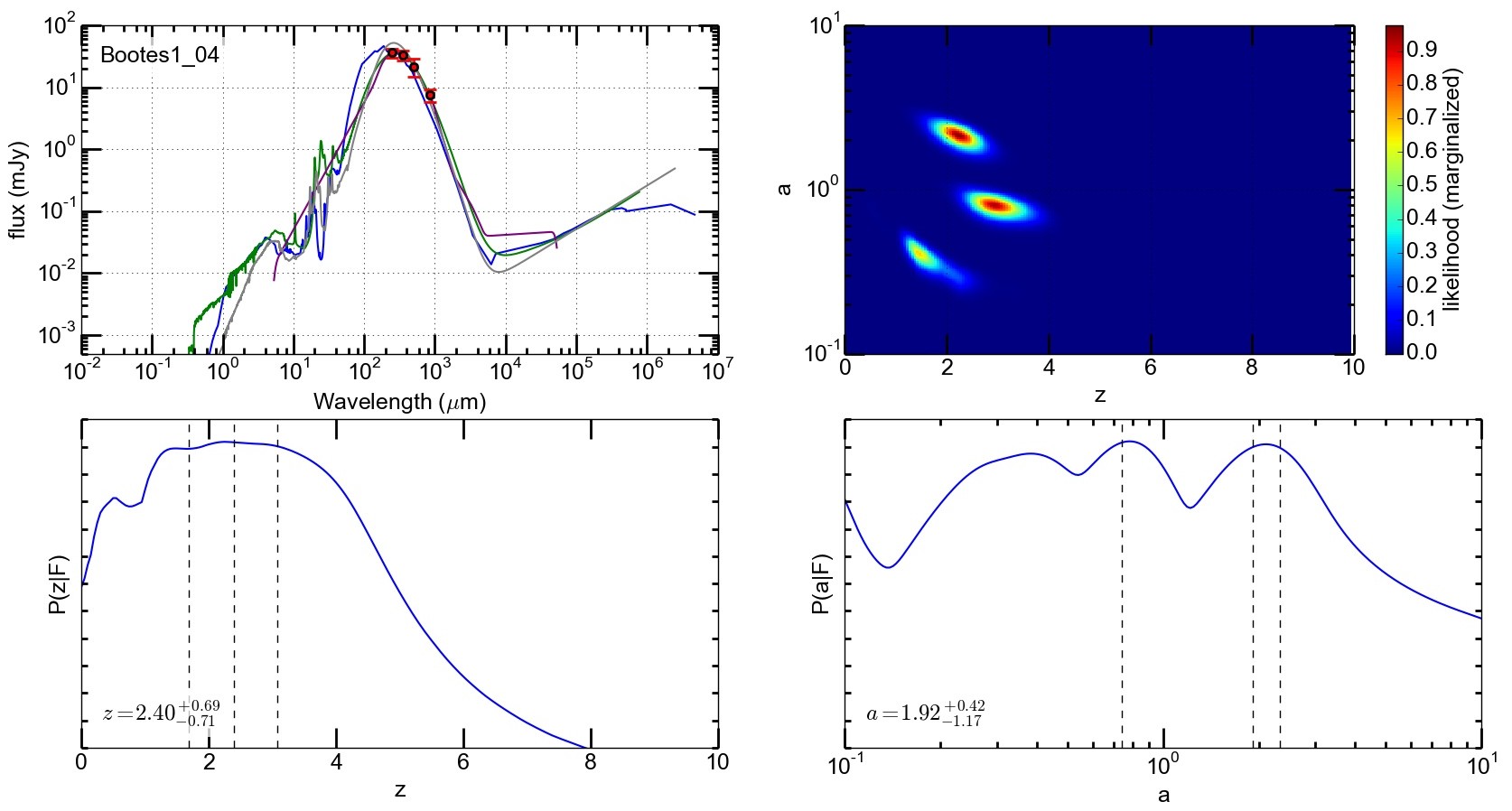}
    \end{subfigure}
    \caption{An example showing the $\chi^{2}$ gridding procedure for one of the sources (Bootes1.04). \textbf{Upper Left:} Redshifted and normalized SEDs of the model templates with the minimum $\chi^{2}$ values. Blue is Arp220, green is ALESS, purple is HFLS3, and grey is Cosmic Eyelash.  The observed flux densities from SCUBA-2 and SPIRE and their associated uncertainties are shown in red. \textbf{Upper Right:} Likelihood values of this source (Bootes1.04) on the parameter space (\textit{z} and \textit{a}). \textbf{Bottom:} Relative probability density functions (PDFs) as functions of \textit{z} and \textit{a}. The vertical black dashed lines represent the 16-th, median, and 84-th quantiles of the PDFs.}\label{Fig5}
\end{figure*}

In order to obtain initial parameter values for the MCMC fitting, we first perform a $\chi^{2}$ gridding search over $z$ and $a$ and the parameters with minimised $\chi^{2}$ are the initial values in the MCMC fitting. Figure \ref{Fig5} shows an example of this $\chi^{2}$ gridding search.  The upper left panel presents the redshifted and normalized SEDs with the observed flux densities and uncertainties. The upper right panel shows likelihood values on the parameter space, \textit{z} and \textit{a}. The lower two panels show the probability density functions (PDFs) with respect to \textit{z} and \textit{a}, respectively.

\begin{figure*}
    \centering
    \begin{subfigure}[b]{0.7\textwidth}
        \includegraphics[width=\columnwidth]{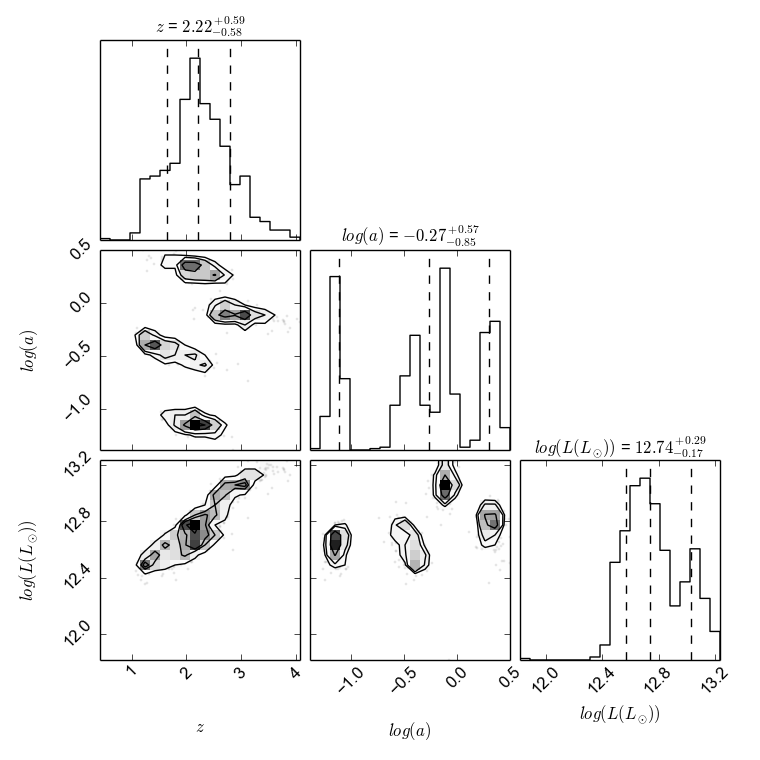}
        \label{Fig7-1}
    \end{subfigure}
    ~ 
    \caption{Posterior distribution of the parameters \textit{z}, \textit{a} and \textit{$L_{\mathrm{IR}}$} for the example source (Bootes1.04) after marginalizing over the four template SEDs. 100 walkers are used, each having 5,000 steps in the constrained parameter ranges $0<z<15$ and $10^{-2}<a<10^{2}$. We choose to implement a burn-in period of 1,000, which are not considered in the fitting. The posterior distributions show 200 randomly selected samples for each template, resulting in 800 samples in total for each source. The black vertical dashed lines indicate the 16th, 50th and 84th percentiles. The best-fit values and uncertainties are given above each posterior distribution.}\label{Fig7}
\end{figure*}

In order to marginalize over the four templates and save computational time, we randomly choose 200 samples from the MCMC chains, giving a total of 800 samples for each source. Figure \ref{Fig7} shows the posterior distribution of the parameters, after maginalization over the templates, of a source in the Bootes1 candidate protocluster (Bootes1.04). Black dashed vertical lines in the posterior distribution show the 16th, 50th and 84th percentiles of the distribution. We use the 50th percentile, or the median, as the best-fit value in this study. The 16th and 84th percentiles of the distribution are given as the range of uncertainty.

Photometric redshift estimates with $z>6$ are regarded as not being robust, due to poor photometry or the redshift-temperature degeneracy. At $z>$6, the SPIRE and SCUBA-2 bands are also no longer near the peak of the modified black-body SED. Hence, we do not include such sources in the discussion of redshift and of infrared luminosity.

In addition to $z$ and $a$, we estimate the infrared luminosity ($L_{\mathrm{IR}}$) and the star-formation rate (SFR) for each of the 800 samples, and thereby determine the posterior distribution, best-fit values and uncertainties for each source. The infrared luminosity is included in the triangle diagram in Fig \ref{Fig7}. The infrared luminosity is estimated by integrating the template SEDs from 8 to 1,000 $\mu$m in the rest frame, given the $z$ and $a$ values for each sample. SFRs are estimated assuming a linear relation between the infrared luminosity \citep{1998ARA&A..36..189K, 2012ARA&A..50..531K}, in which the constant factor of $3.89 \times 10^{-44}$ $M_{\odot} yr^{-1} erg^{-1} s$ and a double-power law stellar initial mass function \citep[IMF,][]{2001MNRAS.322..231K} are used.

Fig \ref{Fig7} and Table \ref{Table2} show that even though the four template SEDs span a variety of different dust temperatures, there are robust estimates in terms of $z$, $L_{\mathrm{IR}}$ and SFR for each source. We can see four distinct peaks in the marginalized posterior distribution of $a$, but these can be attributed to the different normalizations of each template SED and do not affect the conclusions made in this paper.

We notice that there are some sources whose redshifts are poorly estimated, due to the poor photometric data and/or redshift-temperature degeneracy (e.g. Bootes1.07, see Table \ref{Table2}). Nevertheless, they still have robust $L_{\mathrm{IR}}$ and SFR estimates, as suggested in \cite{josh_thesis} (PhD thesis). Although our sources need spectroscopic verifications of their true redshifts, their robust $L_{\mathrm{IR}}$ and SFR estimates indicate that they are likely the most luminous DSFGs ($10^{12} L_{\odot} < L_{\mathrm{IR}} < 10^{13} L_{\odot}$) with the most extreme star-formation activity ($100 \mathrm{M}_{\odot}\mathrm{yr}^{-1} < \mathrm{SFR} < 1,500 \mathrm{M}_{\odot}\mathrm{yr}^{-1}$) within these protoclusters. More photometric data, such as those from millimetre and radio observations of these far-infrared sources would also help to narrow down the photometric redshifts, and on-going observations are being taken or proposed.

\begin{figure}
    \centering
    \begin{subfigure}[b]{0.5\textwidth}
        \includegraphics[width=\columnwidth]{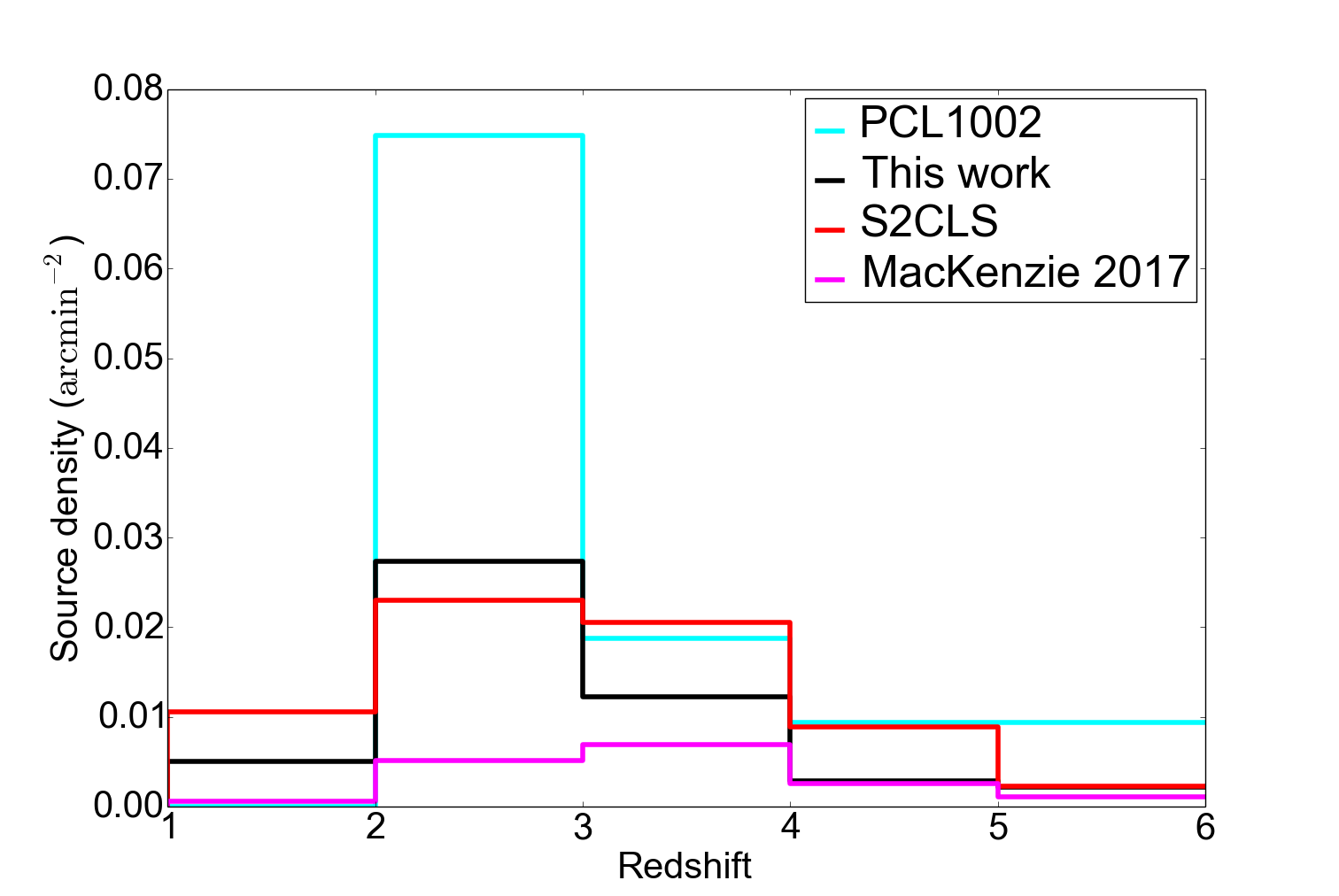}
    \end{subfigure}
    ~ 
    \caption{Photometric redshift distribiton of sources in 13 candidate protocluster fields in this paper (black), a known DSFG-rich protocluster at $z\sim$2, PCL1002 (cyan), 256 sources in the COSMOS field in the S2CLS survey (red), and sources from another \textit{Planck}-based prototcluster candidate survey \citep[magenta,][]{2017MNRAS.468.4006M}. $z > 6$ sources are not included because their photometry no longer covers the peak of the redshifted SED, and their photometric redshifts are considered not robust.}\label{Fig_zall}
\end{figure}

The photometric redshift distribution of all the sources in these 13 candidate protoclusters is shown in Fig \ref{Fig_zall}, in black. The results from \cite{2017MNRAS.468.4006M} are also shown in magenta; this consists of 46 PHz sources \citep{2016A&A...596A.100P} with \textit{Herschel} overdensities, likely in protoclusters. It can be seen that a significant fraction of our sources lie in the range $2 < z < 3$, which is consistent with the expected peak of cosmic star-formation rate density either in protoclusters or in the field \citep{2014MNRAS.439.1193C, 2014ARA&A..52..415M}. This redshift distribution is also consistent with that of the radio/mid-infrared counterparts of 850 $\mu$m sources in the field \citep{2017MNRAS.469..492M}.

We also apply our MCMC $\chi^{2}$-minimisation method to SCUBA-2 maps of a known DSFG-rich protocluster at $z\sim$2, PCL1002 \citep{2015ApJ...808L..33C, 2016ApJ...826..130H}, and 256 sources distributed in the COSMOS field in the S2CLS survey. The original COSMOS map is cropped to the central 1 deg $\times$ 1 deg region in order to exclude the lower sensitivity edges. The resultant photometric redshift distribution is also shown in Fig \ref{Fig_zall}. We note that the photometric redshift distribution peak of our sources in the 13 candidate protoclusters is similar to that of the known $z\sim$2 protocluster PCL1002 and the field sources in the S2CLS/COSMOS field, all showing peaks at $2 < z < 3$. Again this is expected from the studies of the peak of the cosmic star-formation rate density. Sources in \cite{2017MNRAS.468.4006M}, on the other hand, show a peak at $3 < z < 4$, but a significant number of sources lie also within $2 < z < 3$.

We note that PCL1002 has the highest surface density of sources in the range $2 < z < 3$, at approximately 0.074 arcmin$^{-2}$. Our sources in the 13 candidate starbursting protoclusters correspond to approximately 0.027 arcmin$^{-2}$, which is slightly higher than the number of sources in the S2CLS/COSMOS field. SCUBA-2-detected sources in \cite{2017MNRAS.468.4006M} peak at only approximately 0.005 arcmin$^{-2}$.

\section{Discussion}\label{Discussion}

\subsection{Selection Effects}\label{selection_effect}

We compare the results in Gr18 with our results of cumulative number counts shown in Section \ref{number_counts}. In Gr18, 27 candidate DSFG-rich protoclusters were selected based on overdensities of \textit{Herschel}-SPIRE sources at 250, 350 and 500 $\mu$m, whereas in this paper we study number counts at 850 $\mu$m. 

In order to investigate if different methods select different populations of candidate protoclusters, we classify the 13 candidate protoclusters in this paper into four categories: (i) those selected as candidate protoclusters in Gr18, and having an overdensity of 850-$\mu$m sources in this paper; (ii) those selected as candidate protoclusters in Gr18, but not having an overdensity of 850-$\mu$m sources in this paper; (iii) those not selected as candidate protoclusters in Gr18, but having an overdensity of 850-$\mu$m sources in this paper; and (iv) those neither selected as candidate protoclusters in Gr18, nor having an overdensity of 850 $\mu$m sources in this paper. The category for each of the 13 candidate protoclusters in this paper is listed in Table \ref{Table1}.

Bootes1, EGS, NGP4, NGP5, NGP7, and NGP8 are category (i). These candidate protoclusters all have \textit{Herschel} source overdensities at 350 $\mu$m (EGS, NGP5, see Table \ref{Table1}) or 500 $\mu$m (Bootes1, NGP4, NGP7, NGP8), along with their 850-$\mu$m overdensities, suggesting they are high-redshift protoclusters. The colour-colour plot in Fig \ref{Fig1} and the 4-band photometric redshift estimates also suggest that they are bona-fide protoclusters.

Lockman is in category (ii), since it is selected as a candidate protocluster in Gr18 but shows no 850-$\mu$m overdensity. Optical/near-infrared studies on Lockman, using red sequence galaxies, suggest a photometric redshift of $z=2.05 \pm 0.09$ \citep{2014MNRAS.439.1193C}. After cross-matching with \textit{Herschel} sources, the 350$\mu$/850$\mu$m versus 250$\mu$/350$\mu$m colour-colour plot suggests that the two SCUBA-2 sources are at $z>2$ (Fig \ref{Fig1}), and their photometric redshift estimates suggest $z=2.5^{+0.7}_{-0.6}$ and $z=3.1^{+0.7}_{-0.4}$.  However, there are only two 850-$\mu$m sources in this field, and for \textit{Herschel} sources alone it is most overdense at 250 $\mu$m (5.4 $\sigma$, see Table \ref{Table1}). We conclude that Lockman is a lower-redshift ($z<2$) cluster/protocluster, which has an overdensity of 250-$\mu$m sources but such sources become too faint to be detected by SCUBA-2 at 850 $\mu$m. There is also a possibility that our SCUBA-2 maps are not deep enough to detect all of its 850-$\mu$m sources.

G12, NGP2 and NGP3 are in category (iii), showing no significant overdensity in any \textit{Herschel}-SPIRE band (250, 350 or 500 $\mu$m) in Gr18, but they are overdense at 850$\mu$m seen in this paper. According to the colour-colour plot (Fig \ref{Fig1}) and photo-\textit{z} estimates (Table \ref{Table2}), we suspect they are high-redshift ($z>2$) protoclusters, rich in 850-$\mu$m sources. G12 has a spectroscopically confirmed lensed DSFG at $z$=3.26 at the centre of the field \citep{2012ApJ...753..134F, 2013A&A...549A..31H}, so is classified as a lensed source in Gr18, and only has a 2.8 $\sigma$ overdensity at 350 $\mu$m. However, \cite{2016MNRAS.461.1719C} found that G12 is more overdense at submillimetre wavelengths using SCUBA-2 (850 $\mu$m) and LABOCA (870 $\mu$m) observations, consistent with the results shown in this paper. Hence we suspect G12 is a DSFG-rich protocluster where one of the DSFGs happens to be lensed by foreground galaxies and contributes to the bright far-infrared emission seen in \textit{Hershel}-SPIRE and SCUBA-2. NGP2 and NGP3, on the other hand, are probably also high-redshift ($z>2$) protoclusters that are rich in 850-$\mu$m sources.

NGP1, NGP6 and NGP9 are in category (iv), having neither a \textit{Herschel} source overdensity nor 850-$\mu$m overdensity. They might, nevertheless, still be protoclusters, with source overdensities that are too faint to be seen with the observed sensitivity of our \textit{Herschel} or SCUBA-2 observations. Only deeper observations and/or multi-wavelength follow-up observations can determine whether they are protoclusters. Note than even though NGP1 has a 3.2 $\sigma$ overdensity at 500 $\mu$m, it is classified as a lensed far-infrared source in Gr18 due to the bright \textit{Herschel} source in the central region of this field, which is also a confirmed lensed object \citep{2013ApJ...779...25B, 2014ApJ...797..138C, 2016ApJ...829...21T}. Thus NGP1 is not classified as (ii).

\subsection{Infrared Luminosities and Star-formation Rates}\label{LIR}

\begin{figure}
    \centering
    \begin{subfigure}[b]{0.5\textwidth}
        \includegraphics[width=\columnwidth]{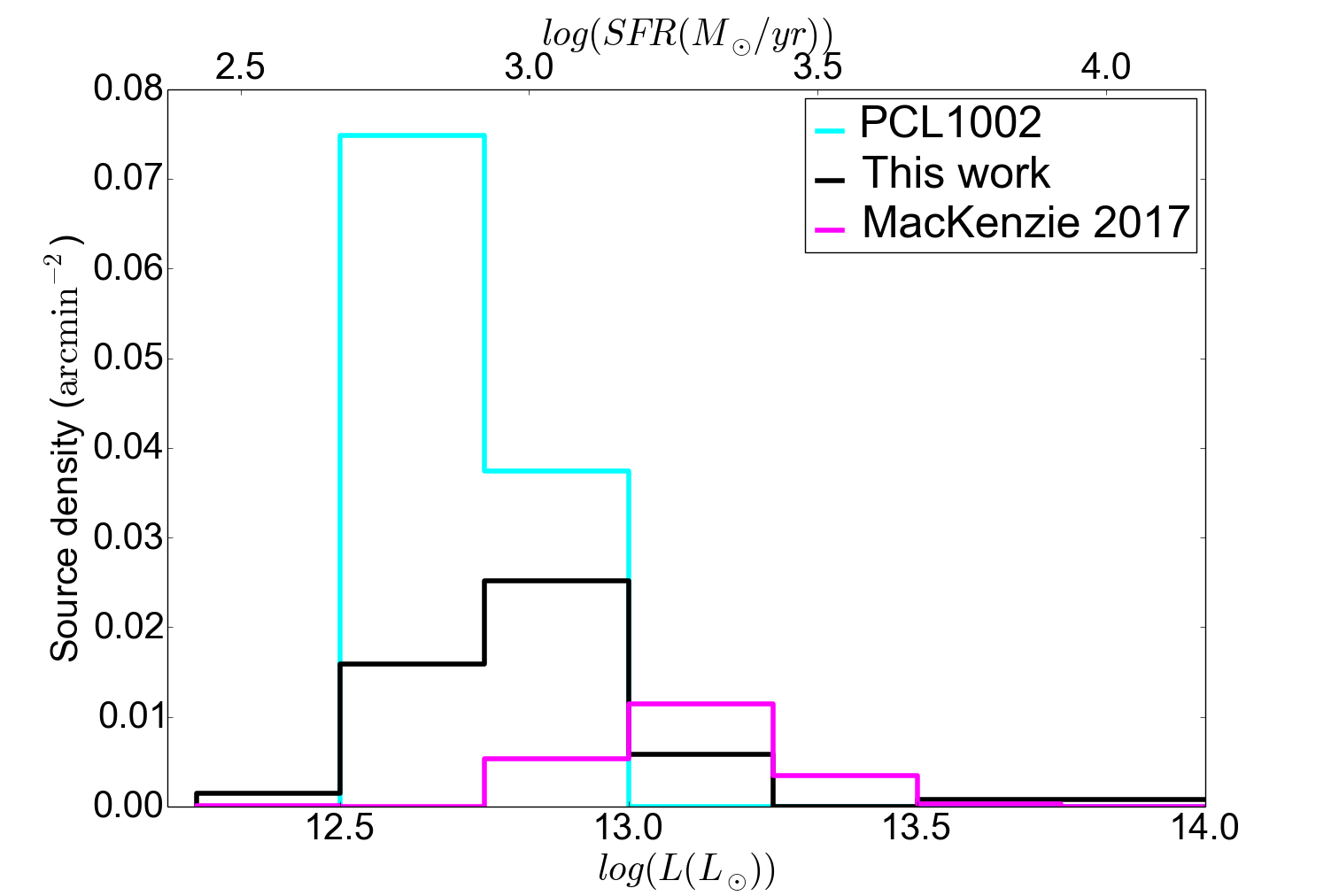}
    \end{subfigure}
    \caption{Distribution of infrared luminosity ($L_{\mathrm{IR}}$) of sources in our 13 candidate protoclusters (black), a known DSFG-rich protocluster at $z\sim$2, PCL1002 (cyan), and sources in candidate protoclusters in \protect\cite{2017MNRAS.468.4006M} (magenta). SFRs are also shown in the upper horizontal axis following a linear relation with the $L_{\mathrm{IR}}$ \citep{1998ARA&A..36..189K}.}
\label{Fig_L}
\end{figure}

As discussed in Section \ref{photo_z}, we estimate the infrared luminosity and SFR for each SCUBA-2 source by integrating the SED from 8 to 1,000 $\mu$m (see Table \ref{Table2}). Fig \ref{Fig_L} shows the $L_{\mathrm{IR}}$ distribution for all the sources in our 13 candidate protoclusters, compared with a known DSFG-rich protocluster at $z\sim$2, PCL1002, and sources in candidate protoclusters from \cite{2017MNRAS.468.4006M}. We have scaled the number of sources to per $arcminute^{2}$ in order to compare fields with different areas.

While PCL1002 has a peak in the bin of $12.5 < \mathrm{log}_{10}(L_{\mathrm{IR}}(\mathrm{L}_{\odot})) < 12.75$ ($3.2 \times 10^{12} < L_{\mathrm{IR}} < 5.6 \times10^{12} L_{\odot}$), our sources in 13 candidate protoclusters show a higher luminosity peak. Nonetheless, a significant number of sources in our candidate protoclusters have infrared luminosities in the same range as the peak in PCL1002. Sources in \cite{2017MNRAS.468.4006M} tend to be more luminous. We conclude that the 850 $\mu$m sources in our candidate protoclusters are as luminous as those of known protoclusters (such as PCL1002), and have representitave infrared luminosities just below $10^{13} L_{\odot}$. We do not totally rule out the possibility of selection effects in these different sets of (candidate) protoclusters (Cheng et al. in prep.), but the study of such selection effect is beyond the scope of this paper.

As discussed in Section \ref{photo_z}, we assume a linear relation between the $L_{\mathrm{IR}}$ and the star-formation rate (SFR) \citep{1998ARA&A..36..189K, 2012ARA&A..50..531K}, with the constant factor being $3.89 \times 10^{-44}$ $M_{\odot} yr^{-1} erg^{-1} s$. In the upper horizontal axis in Fig \ref{Fig_L}, we also quote the SFR values. Similarly the SFRs of the 850-$\mu$m sources in our candidate protoclusters are consistent with the known protocluster PCL1002, and have representative SFRs of $500 \mathrm{M}_{\odot}\mathrm{yr}^{-1} < \mathrm{SFR} < 1,500 \mathrm{M}_{\odot}\mathrm{yr}^{-1}$. There are also a substantial number of sources that have SFR $> 1,000 \mathrm{M}_{\odot}\mathrm{yr}^{-1}$, which might constitute galaxies with extreme star-formation activity or lensed objects.

In confirmed protoclusters at $z>$2, including PCL1002, there are cases of spectroscopically confirmed DSFGs or starbursts having $L_{\mathrm{IR}}$ and SFR values within the range we find here \citep{2014A&A...570A..55D, 2015ApJ...808L..33C, 2015ApJ...815L...8U}. This suggests that even without accurate redshift estimates, approximate infrared luminosities and SFRs can be used to select starbursting members in candidate protoclusters, and this might be applied in general to protoclusters at $z>$2 (Greenslade, 2018, PhD Thesis \footnote{\url{https://spiral.imperial.ac.uk/handle/10044/1/65836}}).

\subsection{Total SFR and protocluster size}\label{totalSFR}

We also estimate the properties of the candidate protoclusters as a whole, by first selecting candidate protocluster member galaxies that have photometric redshifts $z < 6$ and estimating their weighted-mean redshifts. Table \ref{Table_totalSFR} shows the derived properties of the 13 candidate protoclusters studied in this paper. The weighted-mean redshifts are between $1.8 \leq z \leq 3.2$, consistent with them being high-redshift protoclusters.

\begin{table*}
\centering
\scriptsize
\caption{Properties of the 13 candidate protoclusters as derived from individual photometric redshifts and SFRs. \textbf{N}: number of candidate members selected as $z < 6$ sources (see Table \ref{Table2}). \textbf{$z$}: The weighted-mean redshift from the photometric redshift estimates of individual candidate members. Errors are the standard errors on the weighted mean. \textbf{Total SFRs} are estimated by summing up the individual SFRs and \textbf{upper limits} are estimated by fitting the photometry of the \textit{Planck} sources (see text). \textbf{Angular size}, in arcmin, is estimated by the maximum separation of the candidate members. \textbf{Physical size}, in Mpc, is estimated assuming the candidate protocluster is at the weighted-mean redshift. \textbf{Physical volume} is estimated assuming the candidate protoclusters are spherical and the physical size as the diameter. \textbf{SFRD} and its upper limits are estimated by SFR/(physical volume).}
\label{Table_totalSFR}
\begin{tabular}{cccccccc}
\hline Name    & N  & $z$          & \begin{tabular}[c]{@{}c@{}}Total SFR (upper limit)\\ {[}$M_{\odot} yr^{-1}${]}\end{tabular} & \begin{tabular}[c]{@{}c@{}}Angular size\\ {[}arcmin{]}\end{tabular} & \begin{tabular}[c]{@{}c@{}}Physical size\\ {[}Mpc{]}\end{tabular} & \begin{tabular}[c]{@{}c@{}}Physical volume\\ {[}$Mpc^{3}${]}\end{tabular} & \begin{tabular}[c]{@{}c@{}}SFRD (upper limit)\\ {[}$M_{\odot} yr^{-1} Mpc^{-3}${]}\end{tabular} \\ \hline
Bootes1 & 8  & $2.5\pm0.32$ & $7117^{+1821}_{-1394}$ (43946)                                                              & 7.0                                                                 & 3.4                                                               & 14.9                                                                      & 478 (2949)                                                                                      \\
EGS     & 4  & $2.6\pm0.41$ & $5263^{+1475}_{-1121}$ (20150)                                                              & 2.8                                                                 & 1.4                                                               & 1.0                                                                       & 5279 (20150)                                                                                    \\
G12     & 12 & $2.7\pm0.25$ & $22471^{+7311}_{-5365}$ (35145)                                                             & 8.8                                                                 & 4.2                                                               & 29.1                                                                      & 772 (1207)                                                                                      \\
Lockman & 2  & $2.8\pm0.6$  & $1644^{+651}_{-500}$ (15415)                                                                & 8.6                                                                 & 4.1                                                               & 26.6                                                                      & 61 (579)                                                                                        \\
NGP1    & 4  & $1.8\pm0.41$ & $8019^{+6114}_{-6701}$ (24437)                                                              & 4.0                                                                 & 2.0                                                               & 3.3                                                                       & 2399 (7405)                                                                                     \\
NGP2    & 5  & $2.8\pm0.41$ & $6079^{+1949}_{-1166}$ (18019)                                                              & 4.9                                                                 & 2.3                                                               & 4.7                                                                       & 1292 (3833)                                                                                     \\
NGP3    & 1  & $2.6\pm0.88$ & $1885^{+1236}_{-668}$ (18911)                                                               & N/A                                                                 & N/A                                                               & N/A                                                                       & N/A (N/A)                                                                                       \\
NGP4    & 5  & $3.2\pm0.48$ & $6333^{+1672}_{-1237}$ (19083)                                                              & 3.7                                                                 & 1.7                                                               & 1.8                                                                       & 3556 (10601)                                                                                    \\
NGP5    & 5  & $3.0\pm0.4$  & $7052^{+1759}_{-1314}$ (30761)                                                              & 4.0                                                                 & 1.9                                                               & 2.5                                                                       & 2826 (12304)                                                                                    \\
NGP6    & 6  & $2.0\pm0.32$ & $4128^{+1256}_{-1100}$ (24847)                                                              & 5.5                                                                 & 2.8                                                               & 8.2                                                                       & 502 (3030)                                                                                      \\
NGP7    & 10 & $2.9\pm0.3$  & $11527^{+2278}_{-2167}$ (19153)                                                             & 6.3                                                                 & 3.0                                                               & 10.2                                                                      & 1131 (1877)                                                                                     \\
NGP8    & 4  & $2.4\pm0.45$ & $4319^{+1586}_{-1097}$ (21935)                                                              & 6.2                                                                 & 3.0                                                               & 11.1                                                                      & 389 (1976)                                                                                      \\
NGP9    & 3  & $3.2\pm0.65$ & $2806^{+923}_{-866}$ (19567)                                                                & 7.0                                                                 & 3.2                                                               & 12.6                                                                      & 222 (1552)                                                                            \\         \hline
\end{tabular}
\end{table*}

The total SFRs are estimated by summing up the individual SFR of the candidate member galaxies. Since the SFRs are derived using FIR and submm flux densities, the total SFRs may be underestimated due to (partial) dust obscuration. These SFRs should also be considered lower limits since we may miss faint sources in the background that are undetected by SCUBA2. We also estimate the upper limit of the total SFR of each candidate protocluster field by first extracting their \textit{Planck} flux densities at 857, 545, and 352 GHz bands (i.e. 350, 550, and 850 $\mu$m, respectively) from the \textit{Planck} Early Release Compact Source Catalog \citep[ERCSC][]{2011A&A...536A...7P} and the \textit{Planck} Catalogues of Compact Sources \citep[PCCS, PCCS2,][]{2014A&A...571A..28P, 2016A&A...594A..26P}. We then use this 3-band photometry to estimate their SFRs as is performed for the candidate protocluster fields.

The total obscured SFRs of the candidate protoclusters range from approximately 1,600 to 22,500 $M_{\odot} yr^{-1}$ and upper limits for the total SFRs range from approximately 15,400 to 44,000 $M_{\odot} yr^{-1}$. As a reference, the total SFRs of two known protoclusters at $2 < z < 3$ are between 4,900 and 12,500 $M_{\odot} yr^{-1}$ \citep{2019MNRAS.488.1790L}. SFRs of other knwon protoclusters are 6,500 $M_{\odot} yr^{-1}$ for DRC \citep[$z \sim 4$,][]{2018ApJ...856...72O}, $> 1,500 M_{\odot} yr^{-1}$ for AzTEC-3 \citep[$z \sim 5.3$,][]{2011Natur.470..233C}, and $\sim$3,400 $M_{\odot} yr^{-1}$ for CLJ1001 \citep[$z \sim 2.5$,][]{2016ApJ...828...56W}. The four candidate protoclusters studied in \cite{2014MNRAS.439.1193C} have SFRs between 620-11,632 $M_{\odot} yr^{-1}$. Three of them are also studied in this paper (Bootes1, EGS, Lockman). The SFRs of Bootes1 and Lockman are lower than those estimated in \cite{2014MNRAS.439.1193C} whereas the SFR of EGS is higher. We suspect this discrepency is due to the different methods and sources used to estimate the SFRs. In \cite{2014MNRAS.439.1193C} the SFRs are estimated by fitting a modified blackbody with the dust emissivity of $\beta=2$, and the \textit{Herschel} sources with 250, 250 and 500 $\mu$m flux densities are used.

Due to limited source counts, we approximate the size of each candidate protocluste field by the largest angular separation between any pair of candidate member galaxies, and then converting that angular separation to the physical separation at the weighted-mean redshift. The angular separations of the candidate protocluster fields range from 2.8 to 8.8 arcmin, which corresponds to physical separations of 1.4 to 4.2 Mpc. Note that NGP3 has only one candidate member galaxy so we cannot estimate its size. Our candidate protoclusters are likely to span a wide range of physical sizes, which is consistent with the observed diversity of protoclusters in the literature, from the smallest cores of approximately 80 kpc \citep[CLJ1001, $z \sim 2.5$,][]{2016ApJ...828...56W} to large scale filament of approximately 60 Mpc (comoving scale) \citep[SSA22, $z \sim 3.1$,][]{2004AJ....128.2073H}.

Assuming the candidate protocluster fields are spherical, we also estimate their physical volumes by applying the physical separation as the diameter. The SFRDs and their upper limits are also estimated from the total obscured SFRs and their upper limits. The physical volumes of the candidate protoclusters range from 1.0 to 29.1 $Mpc^{3}$. The SFRDs of the candidate protoclusters range from 61 to 5,279 $M_{\odot} yr^{-1} Mpc^{-3}$ with upper limits of 579-20,150 $M_{\odot} yr^{-1} Mpc^{-3}$. 

The SFRDs of our candidate protoclusters are higher than low-redshift or local galaxy clusters, which is consistent with the peak of field SFRD at $2 < z < 3$ (Fig.15 in \cite{2014MNRAS.439.1193C}). These SFRDs indicate that our sample contains some of the most extreme protocluster population, possibly contributing to a large fraction of cosmic SFR at $z > 2$.

\section{Conclusions}\label{Conclusions}

Combining \textit{Planck} and \textit{Herschel} data has revealed a number of candidate protoclusters hosting multiple dusty star-forming galaxies (DSFGs). The abundance of these sources makes these protoclusters some of the most extreme star-forming environments in the Universe. DSFGs are thought to be the progenitors of massive elliptical galaxies residing in the cores of today's massive galaxy clusters. Hence studies of these protocluster galaxies are important in understanding the formation of galaxy clusters and elliptical galaxies. However, only a few high-z protoclusters have been found to date. In this paper we examined 13 candidate protoclusters selected using \textit{Planck} and \textit{Herschel} data and then observed with SCUBA-2.

We calculate the cumulative number counts of SCUBA-2 sources in these candidate protocluster fields. Compared to studies of random fields, nine of our candidate protocluster fields show overdensities of DSFGs.

Combining with the 250-, 350- and 500-$\mu$m flux densities of these SCUBA-2 sources, we estimate their 250$\mu$m/350$\mu$m versus 350$\mu$m/850$\mu$m colours. 11/13 of the candidate protoclusters have colours similar to those of the template SEDs of known starbursts (Arp220, ALESS, HFLS3 and Cosmic Eyelash) redshifted to $z>$2.

We estimate photometric redshifts using the 250-, 350-, 500-, and 850-$\mu$m flux densities using the same template SEDs and a $\chi^{2}$-minimization method. We estimate the infrared luminosity ($L_{\mathrm{IR}}$) of each source by integrating the template SEDs from 8 to 1,000 $\mu$m and determine star-formation rates (SFRs) assuming a linear relation to the infrared luminosity. The redshift distributions of all our sources peak at $2 < z < 3$, which is consistent with the redshift distribution of a known protocluster and the peak of star-formation rate density. We found that the infrared luminosities and SFRs of the sources in our candidate protoclusters are also consistent with those of known protoclusters, and have representative values of $3 \times 10^{12} < L_{\mathrm{IR}} < 10^{13} L_{\odot}$ and $500 \mathrm{M}_{\odot}\mathrm{yr}^{-1} < \mathrm{SFR} < 1,500 \mathrm{M}_{\odot}\mathrm{yr}^{-1}$, respectively. A substantial number of sources in our candidate protoclustes have SFR $> 1,000 \mathrm{M}_{\odot}\mathrm{yr}^{-1}$, suggesting they are starbursting galaxies with some of the most extreme star-forming activity.

We compare our 13 candidate protocluster sample with the 27 candidate protoclusters selected in Gr18. Six of our 13 candidate protoclusters are also selected in Gr18, suggesting they are the most likely bona-fide high-redshift starbursting protoclusters. Lockman is selected by Gr18 but does not have an 850-$\mu$m source overdensity, suggesting it is a lower-redshift protocluster. Three of our 13 candidate protoclusters, which also show overdensities of 850-$\mu$m sources, are not selected in Gr18. One of these has a lensed DSFG in the field (G12) and the other two (NGP2, NGP3) appear to be protoclusters rich in 850-$\mu$m sources at similarly high redshift. Three other candidate protoclusters are not selected in Gr18 and do not have significant overdensities of 850-$\mu$m sources, so are less likely to be true protoclusters, or perhaps the \textit{Herschel} or SCUBA-2 observations are not deep enough to detect associated overdensities of far-infrared or submillimeter sources.

The total obscured SFRs of the candidate protoclusters are estimated and range from approximately 1,600 to 22,500 $M_{\odot} yr^{-1}$ with upper limits of approximately 15,000 to 44,000 $M_{\odot} yr^{-1}$. We also estimate their physical sizes and their SFRDs, concluding our sample contains some of the most extreme protocluster population, possibly contributing a large fraction of cosmic star-formation rate at z$>$2. Future deeper, higher-resolution, multi-wavelength observations (e.g. ALMA, VLT and HST) will help us to understand these early stages of forming galaxy clusters. Those observations include looking for overdensities of optical/near-infrared and mid-infrared sources in these candidate protoclusters, studying the multiplicity rate of the FIR sources, possible weak lensing effects, and potential line-of-sight overlaps of multiple protoclusters in dense DSFGs in the sky. Spectroscopic verifications are also necassary to confirm their protocluster memberships.

\section*{Acknowledgements}
The author appreciates the comments from anonymous referees, M. Zemcov and other collaborators for their insightful comments.

The James Clerk Maxwell Telescope is operated by the East Asian Observatory on behalf of the National Astronomical Observatory of Japan, Academia Sinica Institute of Astronomy and Astrophysics, the Korea Astronomy and Space Science Institute, and the Operation, Maintenance and Upgrading Fund for Astronomical Telescopes and Facility Instruments, budgeted from the Ministry of Finance (MOF) of China and administrated by the Chinese Academy of Sciences (CAS), as well as the National Key R$\&$D Program of China (No. 2017YFA0402700). Additional funding support is provided by the Science and Technology Facilities Council of the United Kingdom and participating universities in the United Kingdom and Canada.

The \textit{Herschel} spacecraft was designed, built, tested, and launched under a contract to ESA managed by the \textit{Herschel}/\textit{Planck} Project team by an industrial consortium under the overall responsibility of the prime contractor Thales Alenia Space (Cannes), and including Astrium (Friedrichshafen) responsible for the payload module and for system testing at spacecraft level, Thales Alenia Space (Turin) responsible for the service module, and Astrium (Toulouse) responsible for the telescope, with in excess of a hundred subcontractors.

SPIRE has been developed by a consortium of institutes led by Cardiff University (UK) and including: Univ. Lethbridge (Canada); NAOC (China); CEA, and LAM (France); IFSI, Univ. Padua (Italy); IAC (Spain); Stockholm Observatory (Sweden); Imperial College London, RAL, UCL-MSSL, UKATC, and Univ. Sussex (UK); and Caltech, JPL, NHSC, and Univ. Colorado (USA). This development has been supported by national funding agencies: CSA (Canada); NAOC (China); CEA, CNES, CNRS (France); ASI (Italy); MCINN (Spain); SNSB (Sweden); STFC, UKSA (UK); and NASA (USA).

This research has made use of data from HerMES project (\url{http://hermes.sussex.ac.uk/}). HerMES is a Herschel Key Programme utilising Guaranteed Time from the SPIRE instrument team, ESAC scientists and a mission scientist.
The HerMES data was accessed through the Herschel Database in Marseille (HeDaM - \url{http://hedam.lam.fr}) operated by CeSAM and hosted by the Laboratoire d'Astrophysique de Marseille.

GDZ gratefully acknowledges financial support from ASI/INAF agreement n.~2014-024-R.1 for the {\it Planck} LFI Activity of Phase E2 and from the ASI/Physics Department of the university of Roma--Tor Vergata agreement n. 2016-24-H.0

JGN acknowledges financial support from the I+D 2015 project AYA2015- 65887-P (MINECO/FEDER) and from the Spanish MINECO for a ``Ramon y Cajal" fellowship (RYC-2013-13256).

E.I.\ acknowledges partial support from FONDECYT through grant N$^\circ$\,1171710.

M.J.M.~acknowledges the support of the National Science Centre, Poland through the POLONEZ grant 2015/19/P/ST9/04010 and SONATA BIS grant 2018/30/E/ST9/00208; this project has received funding from the European Union's Horizon 2020 research and innovation programme under the Marie Sk{\l}odowska-Curie grant agreement No. 665778.

D.R. acknowledges support from the National Science Foundation under grant number AST-1614213.

H.D. acknowledges financial support from the Spanish Ministry of Science, Innovation and Universities (MICIU) under the 2014 Ram$\acute{o}$n y Cajal program RYC-2014-15686 and AYA2017-84061-P, the later one co-financed by FEDER (European Regional Development Funds).




\bibliographystyle{mnras}
\bibliography{SCUBA2_protocluster_tcheng_190917} 




\appendix

\section{SCUBA-2 850 $\mu$m flux maps}

Fig.\ref{Fig2} shows SCUBA-2 850-$\mu$m flux maps of the 13 candidate prorocluster fields discussed in this paper.

\begin{figure*}
    \centering
    \begin{subfigure}[b]{0.65\textwidth}
        \includegraphics[width=\columnwidth]{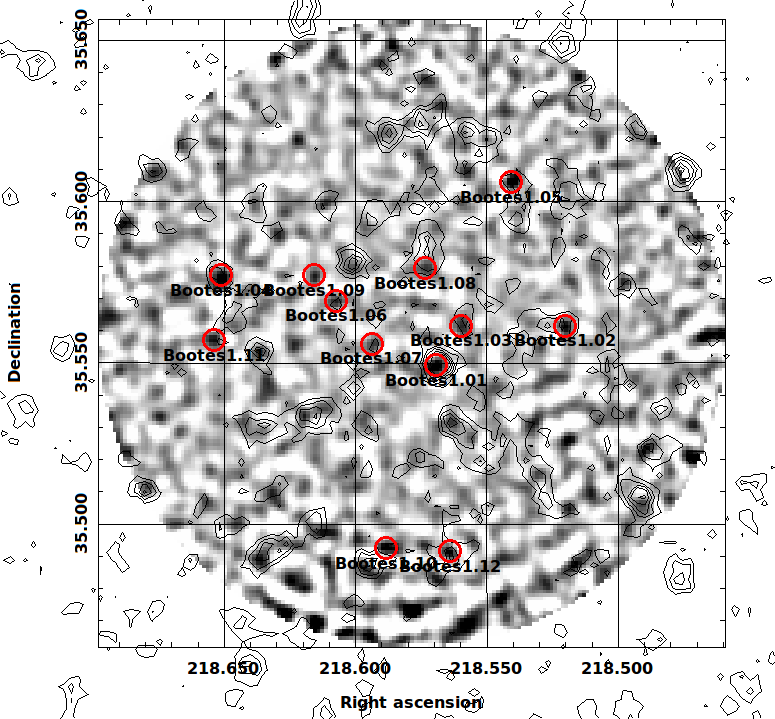}
        \caption{Bootes1}
        \label{Fig2-1}
    \end{subfigure}
    \begin{subfigure}[b]{0.65\textwidth}
        \includegraphics[width=\columnwidth]{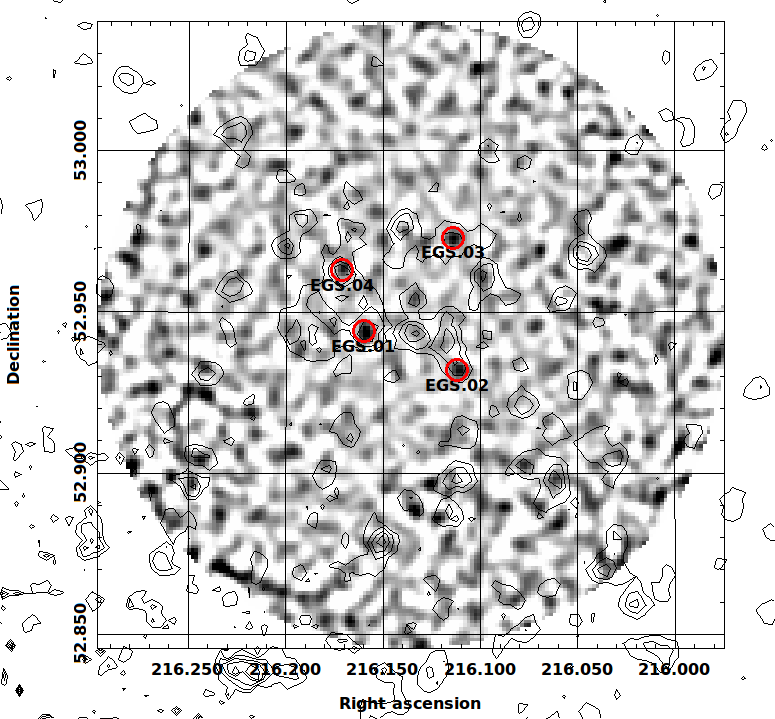}
        \caption{EGS}
        \label{Fig2-2}
    \end{subfigure}
    \caption{SCUBA-2 850-$\mu$m flux maps of the 13 candidate prorocluster fields. Detected sources ($>$3.5 $\sigma$) are shown as red circles, labelled with their names. Black contours are \textit{Herschel} 350 $\mu$m flux densities, with levels of 10, 20, 30, 40 and 50 mJy.}\label{Fig2}
\end{figure*}

\begin{figure*}
    \ContinuedFloat 
    \centering
    \begin{subfigure}[b]{0.65\textwidth}
        \includegraphics[width=\columnwidth]{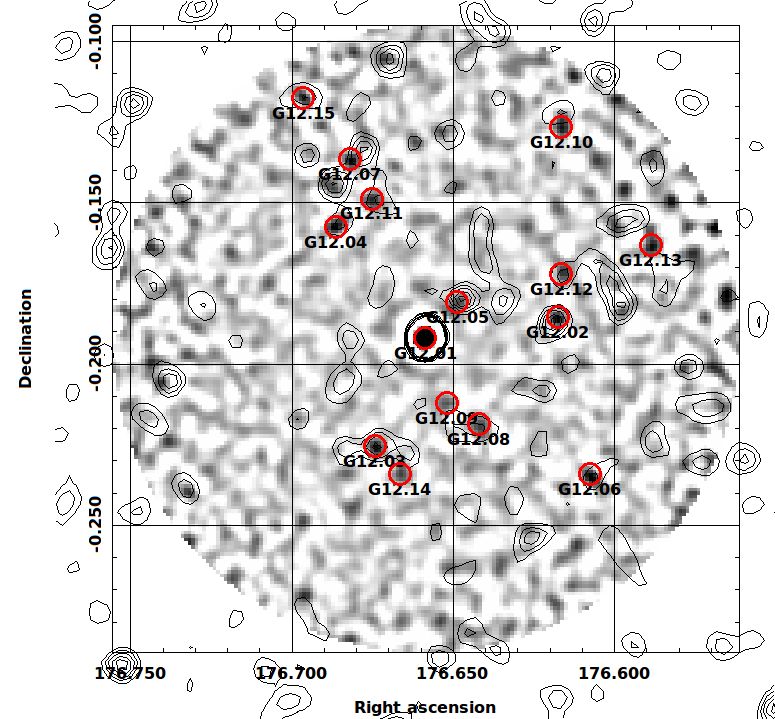}
        \caption{G12}
        \label{Fig2-3}
    \end{subfigure}
        \begin{subfigure}[b]{0.65\textwidth}
        \includegraphics[width=\columnwidth]{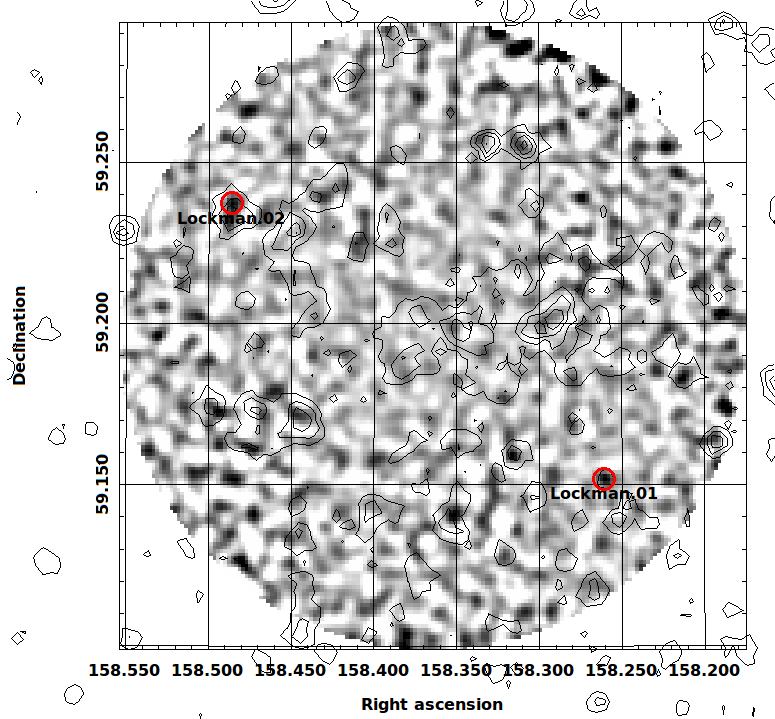}
        \caption{Lockman}
        \label{Fig2-4}
    \end{subfigure}
    \caption{Continued.}\label{Fig2_2}
\end{figure*}

\begin{figure*}
    \ContinuedFloat 
    \centering
        \begin{subfigure}[b]{0.65\textwidth}
        \includegraphics[width=\columnwidth]{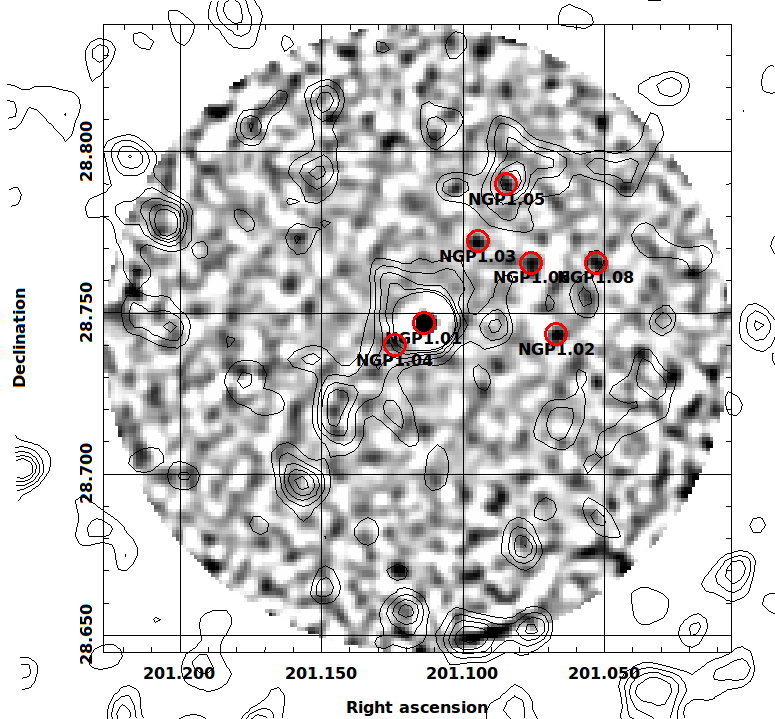}
        \caption{NGP1}
        \label{Fig2-5}
    \end{subfigure}
        \begin{subfigure}[b]{0.65\textwidth}
        \includegraphics[width=\columnwidth]{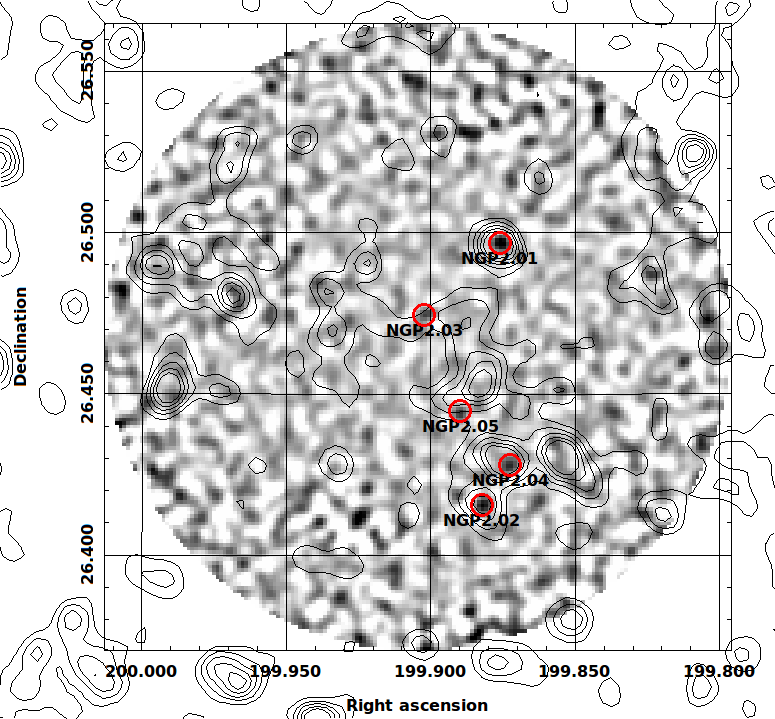}
        \caption{NGP2}
        \label{Fig2-6}
    \end{subfigure}
    \caption{Continued.}\label{Fig2_3}
\end{figure*}

\begin{figure*}
    \ContinuedFloat 
    \centering
    \begin{subfigure}[b]{0.65\textwidth}
        \includegraphics[width=\columnwidth]{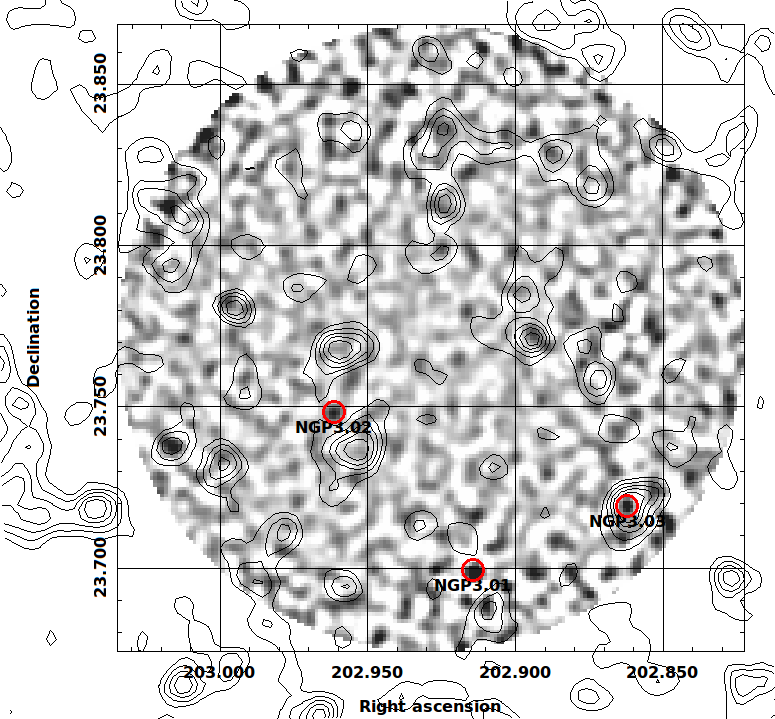}
        \caption{NGP3}
        \label{Fig2-7}
    \end{subfigure}
    ~
    \begin{subfigure}[b]{0.65\textwidth}
        \includegraphics[width=\columnwidth]{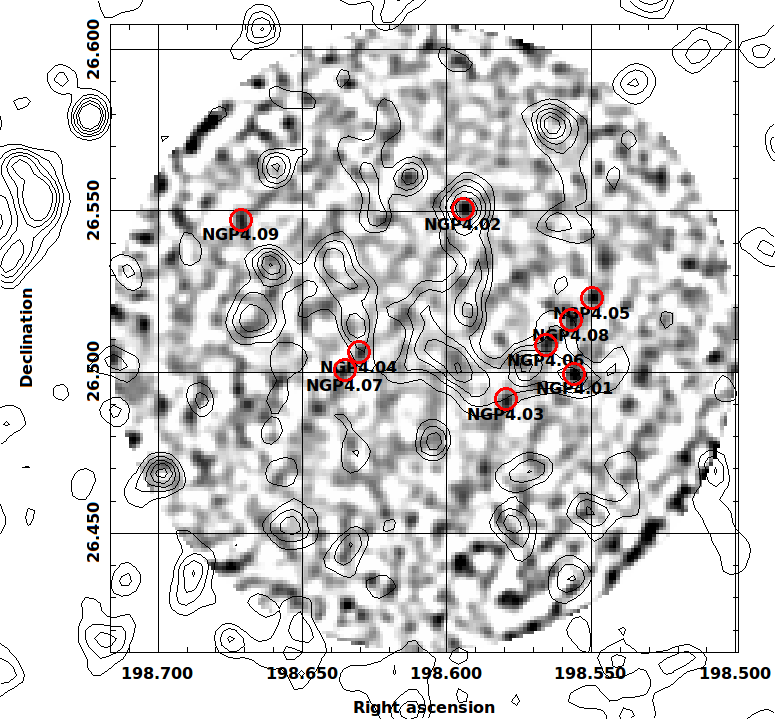}
        \caption{NGP4}
        \label{Fig2-8}
    \end{subfigure}
    \caption{Continued.}\label{Fig2_4}
\end{figure*}

\begin{figure*}
    \ContinuedFloat 
    \centering
    \begin{subfigure}[b]{0.65\textwidth}
        \includegraphics[width=\columnwidth]{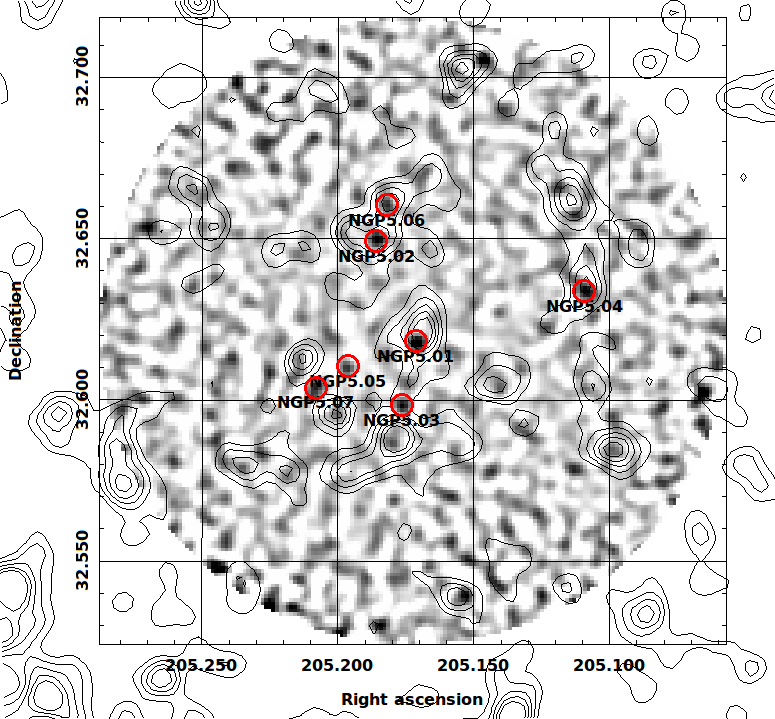}
        \caption{NGP5}
        \label{Fig2-9}
    \end{subfigure}
    ~
    \begin{subfigure}[b]{0.65\textwidth}
        \includegraphics[width=\columnwidth]{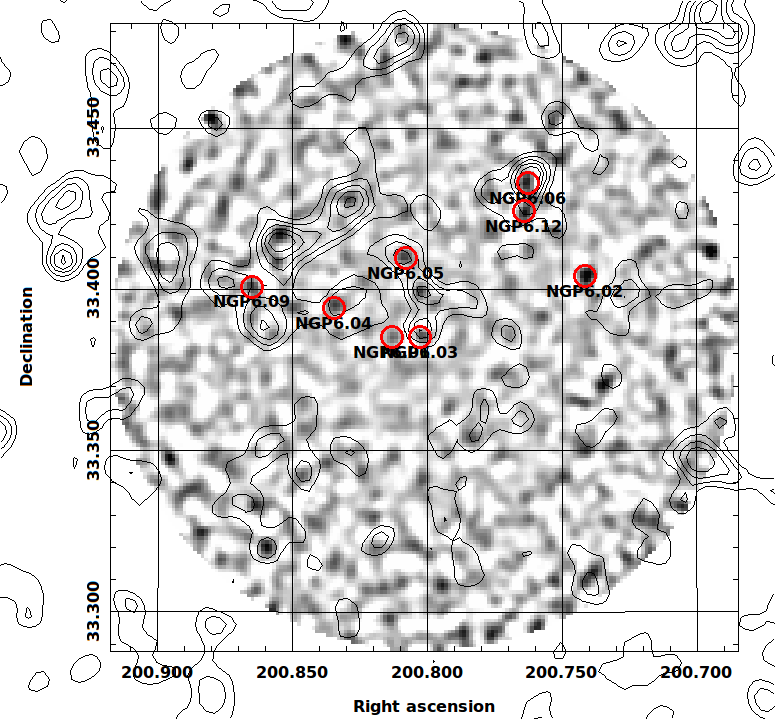}
        \caption{NGP6}
        \label{Fig2-10}
    \end{subfigure}
    \caption{Continued.}\label{Fig2_5}
\end{figure*}

\begin{figure*}
    \ContinuedFloat 
    \centering
    \begin{subfigure}[b]{0.65\textwidth}
        \includegraphics[width=\columnwidth]{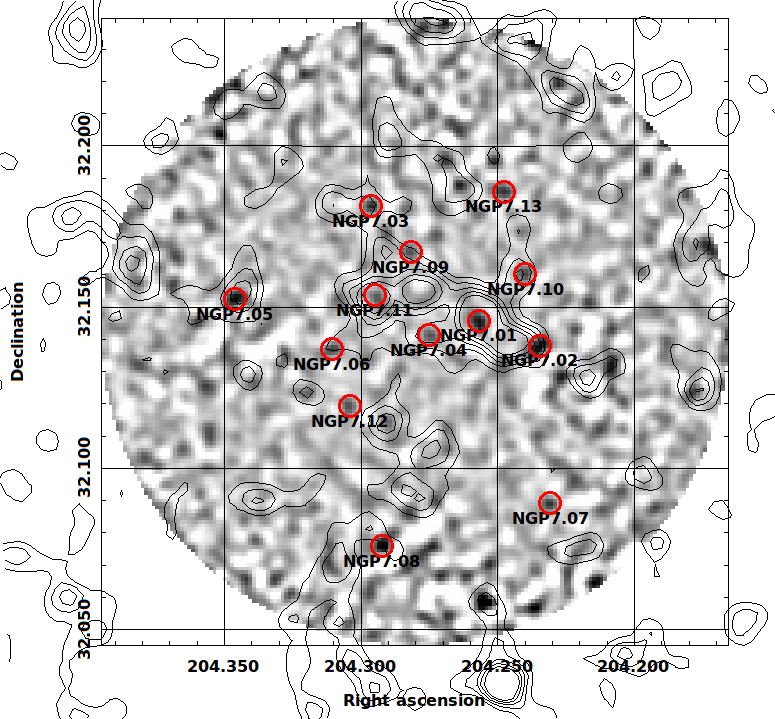}
        \caption{NGP7}
        \label{Fig2-11}
    \end{subfigure}
    ~
    \begin{subfigure}[b]{0.65\textwidth}
        \includegraphics[width=\columnwidth]{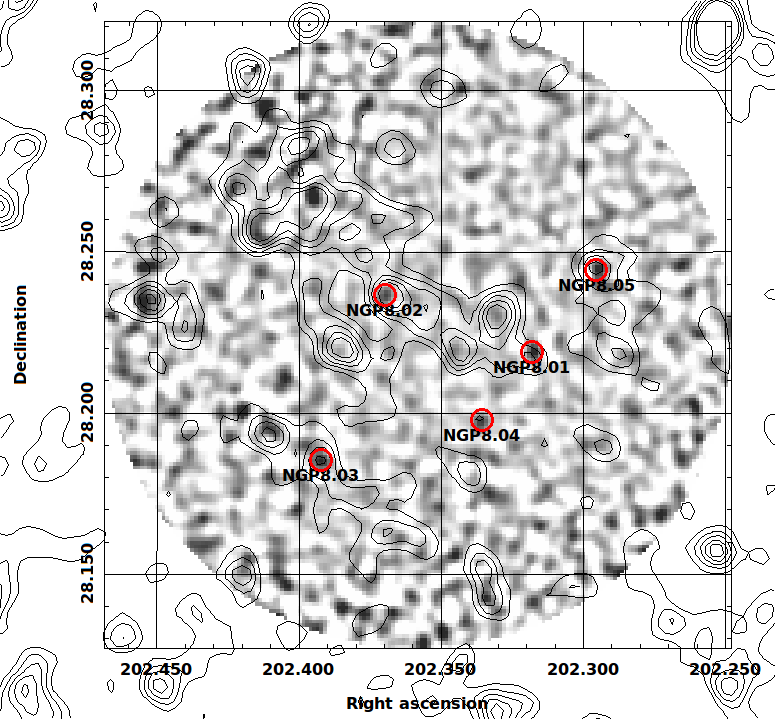}
        \caption{NGP8}
        \label{Fig2-12}
    \end{subfigure}
    \caption{Continued.}\label{Fig2_6}
\end{figure*}

\begin{figure*}
    \ContinuedFloat 
    \centering
    \begin{subfigure}[b]{0.65\textwidth}
        \includegraphics[width=\columnwidth]{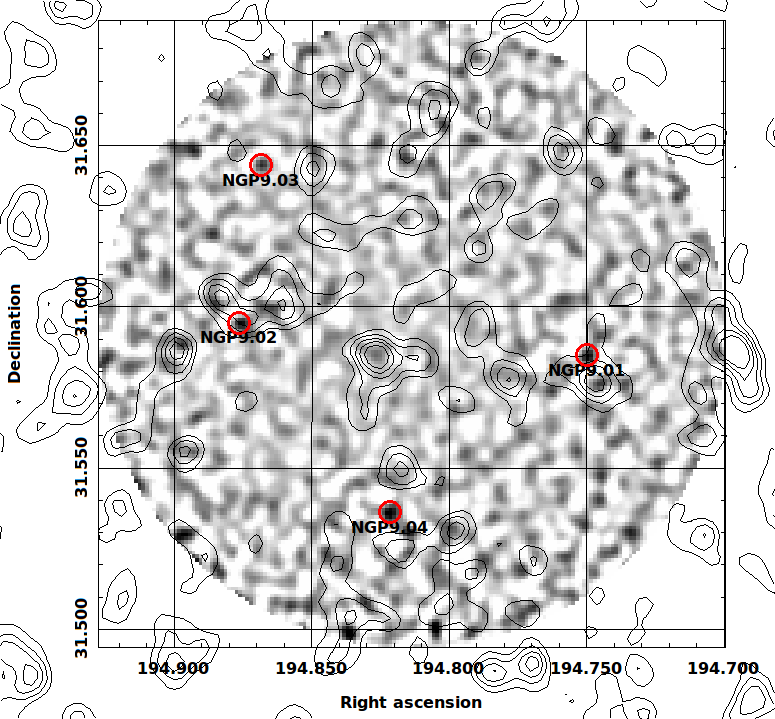}
        \caption{NGP9}
        \label{Fig2-13}
    \end{subfigure}
    \caption{Continued.}\label{Fig2_7}
\end{figure*}

\section{Source Catalogues}

Table \ref{Table2} shows the SCUBA-2 source catalogue of the 13 candidate protoclusters discussed in this paper.

\begin{landscape}

\begin{table}
\centering
\scriptsize
\caption{Catalogue of SCUBA-2 sources in the 13 candidate protocluster fields. \textbf{Column 4}: signal-to-noise ratio at 850 $\mu$m. \textbf{Column 5}: $S_{850}$($\sigma_{850}$) de-boosted flux densities and uncertainties of the detected SCUBA-2 sources at 850 $\mu$m, after the confusion noise of 0.7 mJy and the 5 percent calibraton error are added in quadrature. \textbf{Column 6--8}: $S_{250}$, $S_{350}$, $S_{500}$ ($\sigma_{250}$, $\sigma_{350}$, $\sigma_{500}$), SPIRE flux densities and uncertainties; if determined from the SPIRE maps (class = map), the uncertainties are quoted as the confusion limits at the respective bands. \textbf{Column 9}: if class=$cat$, the SPIRE flux densities and uncertainties are quoted from the \textit{Herschel} catalogues, and $cat*$ means that the \textit{Herschel} counterpart has a separation between 9'' and 18'', which has a higher rate of being a spurious match as discussed in \ref{colours}; if class=$map$, the SPIRE flux densities are quoted in the maps at the positions of the SCUBA-2 sources. \textbf{Column 10--12}: $z$, $log(L_{\mathrm{IR}})$ and $log(\mathrm{SFR})$, marginalized photometric redshift, (log of) infrared luminosity and (log of) star-formation rates, respectively. These are found by fitting the 250, 350, 500 and 850 flux densities, using the MCMC method. The best-fit is gven as the median and errors are quoted as 16th and 84th percentiles. ($\#$): sources in NGP3 where the completeness is below 50 percent.}
\label{Table2}
\begin{tabular}{cccccccccccc}
\hline
\multicolumn{1}{|c|}{Name} & \multicolumn{1}{c|}{\begin{tabular}[c]{@{}c@{}}RA\\ (J2000)\end{tabular}} & \multicolumn{1}{c|}{\begin{tabular}[c]{@{}c@{}}Dec\\ (J2000)\end{tabular}} & \multicolumn{1}{c|}{\begin{tabular}[c]{@{}c@{}}S/N \\ (at 850$\mu$m)\end{tabular}} & \multicolumn{1}{c|}{\begin{tabular}[c]{@{}c@{}}$S_{850}$ ($\sigma_{850}$)\\ {[}mJy{]}\end{tabular}} & \multicolumn{1}{c|}{\begin{tabular}[c]{@{}c@{}}$S_{250}$ ($\sigma_{250}$)\\ {[}mJy{]}\end{tabular}} & \multicolumn{1}{c|}{\begin{tabular}[c]{@{}c@{}}$S_{350}$ ($\sigma_{350}$)\\ {[}mJy{]}\end{tabular}} & \multicolumn{1}{c|}{\begin{tabular}[c]{@{}c@{}}$S_{500}$ ($\sigma_{500}$)\\ {[}mJy{]}\end{tabular}} & \multicolumn{1}{c|}{Class} & \multicolumn{1}{c|}{Redshift} & \multicolumn{1}{c|}{$log(L_{\mathrm{IR}}[L_{\odot}])$} & \multicolumn{1}{c|}{$log(SFR[M_{\odot}yr^{-1}])$} \\ \hline
\multicolumn{12}{|c|}{Bootes1}                                                                                                                                                                                                                                                                                                                                                                                                                                                                                                                                                                                                                                                                                                                                                                                                                                             \\ \hline
Bootes1.01                 & 218.56919                                                                 & 35.54917                                                                   & 7.8                                                                                & 9.9 (1.4)                                                                                           & 132.6 (6.3)                                                                                         & 96.0 (6.5)                                                                                          & 65.9 (6.9)                                                                                          & cat                        & $1.43^{+0.58}_{-0.8}$         & $12.94^{+0.41}_{-0.35}$                                & $3.11^{+0.41}_{-0.35}$                            \\
Bootes1.02                 & 218.52002                                                                 & 35.56137                                                                   & 5.1                                                                                & 7.4 (1.5)                                                                                           & 23.8 (6.3)                                                                                          & 32.7 (6.5)                                                                                          & 32.2 (7.1)                                                                                          & cat                        & $2.72^{+0.77}_{-0.44}$        & $12.83^{+0.28}_{-0.15}$                                & $3.0^{+0.28}_{-0.15}$                             \\
Bootes1.03                 & 218.55963                                                                 & 35.56139                                                                   & 4.6                                                                                & 5.3 (1.2)                                                                                           & 9.7 (6.3)                                                                                           & 18.7 (6.6)                                                                                          & 15.7 (7.2)                                                                                          & map                        & $3.41^{+0.88}_{-0.76}$        & $12.67^{+0.22}_{-0.19}$                                & $2.85^{+0.22}_{-0.19}$                            \\
Bootes1.04                 & 218.65115                                                                 & 35.57692                                                                   & 4.5                                                                                & 7.5 (1.7)                                                                                           & 36.2 (6.3)                                                                                          & 33.2 (6.5)                                                                                          & 21.7 (6.9)                                                                                          & cat                        & $2.22^{+0.59}_{-0.58}$        & $12.74^{+0.29}_{-0.17}$                                & $2.91^{+0.29}_{-0.17}$                            \\
Bootes1.05                 & 218.54048                                                                 & 35.60583                                                                   & 4.5                                                                                & 9.0 (2.1)                                                                                           & 17.9 (6.3)                                                                                          & 18.4 (6.6)                                                                                          & 27.9 (7.2)                                                                                          & map                        & $3.54^{+0.9}_{-0.58}$         & $12.91^{+0.2}_{-0.18}$                                 & $3.09^{+0.2}_{-0.18}$                             \\
Bootes1.06                 & 218.60744                                                                 & 35.56916                                                                   & 4.4                                                                                & 5.4 (1.3)                                                                                           & 10.1 (6.3)                                                                                          & 11.9 (6.6)                                                                                          & 9.0 (7.2)                                                                                           & map                        & $4.09^{+2.31}_{-1.21}$        & $12.68^{+0.23}_{-0.22}$                                & $2.85^{+0.23}_{-0.22}$                            \\
Bootes1.07                 & 218.59377                                                                 & 35.55583                                                                   & 4.3                                                                                & 4.6 (1.1)                                                                                           & -9.4 (6.3)                                                                                          & 16.7 (6.6)                                                                                          & 4.1 (7.2)                                                                                           & map                        & $8.09^{+4.66}_{-3.34}$        & $12.82^{+0.3}_{-0.29}$                                 & $2.99^{+0.3}_{-0.29}$                             \\
Bootes1.08                 & 218.57329                                                                 & 35.57917                                                                   & 4.1                                                                                & 4.6 (1.1)                                                                                           & 25.8 (6.3)                                                                                          & 25.1 (6.6)                                                                                          & 26.9 (7.3)                                                                                          & cat                        & $2.15^{+0.78}_{-0.71}$        & $12.58^{+0.33}_{-0.19}$                                & $2.75^{+0.33}_{-0.19}$                            \\
Bootes1.09                 & 218.61564                                                                 & 35.57694                                                                   & 4.1                                                                                & 5.4 (1.4)                                                                                           & -3.1 (6.3)                                                                                          & 4.1 (6.6)                                                                                           & 3.3 (7.2)                                                                                           & map                        & $10.18^{+3.29}_{-3.78}$       & $12.97^{+0.26}_{-0.28}$                                & $3.15^{+0.26}_{-0.28}$                            \\
Bootes1.10                 & 218.58830                                                                 & 35.49250                                                                   & 4.0                                                                                & 7.1 (1.8)                                                                                           & 3.0 (6.3)                                                                                           & 1.8 (6.6)                                                                                           & -2.6 (7.2)                                                                                          & map                        & $11.14^{+2.89}_{-3.65}$       & $13.12^{+0.25}_{-0.3}$                                 & $3.29^{+0.25}_{-0.3}$                             \\
Bootes1.11                 & 218.65387                                                                 & 35.55692                                                                   & 4.0                                                                                & 5.7 (1.5)                                                                                           & 2.8 (6.3)                                                                                           & 2.1 (6.6)                                                                                           & -0.6 (7.2)                                                                                          & map                        & $10.54^{+3.13}_{-4.01}$       & $13.01^{+0.26}_{-0.32}$                                & $3.18^{+0.26}_{-0.32}$                            \\
Bootes1.12                 & 218.56374                                                                 & 35.49139                                                                   & 4.0                                                                                & 6.2 (1.6)                                                                                           & 23.1 (6.3)                                                                                          & 32.6 (6.6)                                                                                          & 24.5 (7.2)                                                                                          & cat                        & $2.54^{+0.86}_{-0.51}$        & $12.74^{+0.3}_{-0.16}$                                 & $2.91^{+0.3}_{-0.16}$                             \\ \hline
\multicolumn{12}{|c|}{EGS1}                                                                                                                                                                                                                                                                                                                                                                                                                                                                                                                                                                                                                                                                                                                                                                                                                                                \\ \hline
EGS.01                     & 216.15980                                                                 & 52.94400                                                                   & 6.4                                                                                & 12.3 (2.0)                                                                                          & 43.5 (3.5)                                                                                          & 40.4 (4.1)                                                                                          & 32.2 (4.6)                                                                                          & cat                        & $2.48^{+0.62}_{-0.39}$        & $12.96^{+0.23}_{-0.18}$                                & $3.13^{+0.23}_{-0.18}$                            \\
EGS.02                     & 216.11187                                                                 & 52.93178                                                                   & 4.5                                                                                & 8.3 (1.9)                                                                                           & 24.5 (3.5)                                                                                          & 30.4 (4.1)                                                                                          & 25.9 (4.4)                                                                                          & cat                        & $2.94^{+0.73}_{-0.42}$        & $12.9^{+0.22}_{-0.2}$                                  & $3.08^{+0.22}_{-0.2}$                             \\
EGS.03                     & 216.11369                                                                 & 52.97287                                                                   & 4.5                                                                                & 8.8 (2.0)                                                                                           & 31.2 (3.5)                                                                                          & 39.2 (4.1)                                                                                          & 36.7 (4.4)                                                                                          & cat                        & $2.92^{+0.68}_{-0.31}$        & $13.03^{+0.2}_{-0.19}$                                 & $3.2^{+0.2}_{-0.19}$                              \\
EGS.04                     & 216.17088                                                                 & 52.96288                                                                   & 4.0                                                                                & 7.3 (1.9)                                                                                           & 59.8 (3.5)                                                                                          & 57.9 (4.1)                                                                                          & 38.9 (4.6)                                                                                          & cat                        & $2.01^{+0.64}_{-0.68}$        & $12.88^{+0.33}_{-0.15}$                                & $3.05^{+0.33}_{-0.15}$                            \\ \hline
\multicolumn{12}{|c|}{G12}                                                                                                                                                                                                                                                                                                                                                                                                                                                                                                                                                                                                                                                                                                                                                                                                                                                 \\ \hline
G12.01                     & 176.65833                                                                 & -0.19222                                                                   & 59.6                                                                               & 79.8 (4.2)                                                                                          & 316.0 (6.6)                                                                                         & 357.9 (7.4)                                                                                         & 291.8 (7.7)                                                                                         & cat                        & $2.6^{+0.83}_{-0.15}$         & $13.92^{+0.25}_{-0.18}$                                & $4.1^{+0.25}_{-0.18}$                             \\
G12.02                     & 176.61722                                                                 & -0.18556                                                                   & 6.0                                                                                & 9.3 (1.6)                                                                                           & 49.0 (6.5)                                                                                          & 55.1 (7.4)                                                                                          & 27.5 (7.8)                                                                                          & cat                        & $2.11^{+0.62}_{-0.66}$        & $12.84^{+0.32}_{-0.15}$                                & $3.02^{+0.32}_{-0.15}$                            \\
G12.03                     & 176.67389                                                                 & -0.22556                                                                   & 5.8                                                                                & 8.8 (1.6)                                                                                           & 48.7 (6.6)                                                                                          & 43.7 (7.5)                                                                                          & 40.0 (7.7)                                                                                          & cat                        & $2.2^{+0.79}_{-0.82}$         & $12.86^{+0.34}_{-0.19}$                                & $3.03^{+0.34}_{-0.19}$                            \\
G12.04                     & 176.68611                                                                 & -0.15778                                                                   & 5.3                                                                                & 9.0 (1.7)                                                                                           & 15.9 (7.3)                                                                                          & 20.0 (8.1)                                                                                          & 22.0 (8.6)                                                                                          & map                        & $3.57^{+0.93}_{-0.55}$        & $12.88^{+0.21}_{-0.16}$                                & $3.05^{+0.21}_{-0.16}$                            \\
G12.05                     & 176.64833                                                                 & -0.18111                                                                   & 5.0                                                                                & 5.8 (1.2)                                                                                           & 54.9 (6.5)                                                                                          & 55.4 (7.5)                                                                                          & 33.9 (7.8)                                                                                          & cat                        & $1.7^{+0.49}_{-1.07}$         & $12.71^{+0.34}_{-0.43}$                                & $2.88^{+0.34}_{-0.43}$                            \\
G12.06                     & 176.60722                                                                 & -0.23444                                                                   & 5.0                                                                                & 8.8 (1.8)                                                                                           & 13.6 (7.3)                                                                                          & 19.5 (8.1)                                                                                          & 15.8 (8.6)                                                                                          & map                        & $3.73^{+1.01}_{-0.64}$        & $12.87^{+0.21}_{-0.2}$                                 & $3.04^{+0.21}_{-0.2}$                             \\
G12.07                     & 176.68167                                                                 & -0.13667                                                                   & 4.7                                                                                & 8.5 (1.9)                                                                                           & 7.5 (7.3)                                                                                           & 21.9 (8.1)                                                                                          & 22.3 (8.6)                                                                                          & map                        & $3.99^{+0.93}_{-0.64}$        & $12.9^{+0.18}_{-0.17}$                                 & $3.07^{+0.18}_{-0.17}$                            \\
G12.08                     & 176.64167                                                                 & -0.21889                                                                   & 4.4                                                                                & 5.9 (1.4)                                                                                           & 14.8 (7.3)                                                                                          & 21.6 (8.1)                                                                                          & 20.7 (8.6)                                                                                          & map                        & $3.09^{+0.89}_{-0.6}$         & $12.72^{+0.24}_{-0.18}$                                & $2.89^{+0.24}_{-0.18}$                            \\
G12.09                     & 176.65167                                                                 & -0.21222                                                                   & 4.4                                                                                & 5.1 (1.2)                                                                                           & 1.8 (7.3)                                                                                           & 4.6 (8.1)                                                                                           & 12.9 (8.6)                                                                                          & map                        & $7.03^{+4.54}_{-2.53}$        & $12.82^{+0.26}_{-0.25}$                                & $2.99^{+0.26}_{-0.25}$                            \\
G12.10                     & 176.61611                                                                 & -0.12667                                                                   & 4.3                                                                                & 7.5 (1.8)                                                                                           & 25.0 (6.1)                                                                                          & 20.1 (7.4)                                                                                          & 4.8 (7.7)                                                                                           & cat*                       & $2.57^{+0.88}_{-0.71}$        & $12.67^{+0.25}_{-0.2}$                                 & $2.84^{+0.25}_{-0.2}$                             \\
G12.11                     & 176.67500                                                                 & -0.14889                                                                   & 4.2                                                                                & 6.6 (1.6)                                                                                           & 24.5 (6.0)                                                                                          & 24.1 (7.4)                                                                                          & 16.6 (7.7)                                                                                          & cat*                       & $2.61^{+0.76}_{-0.61}$        & $12.7^{+0.25}_{-0.18}$                                 & $2.88^{+0.25}_{-0.18}$                            \\
G12.12                     & 176.61611                                                                 & -0.17222                                                                   & 4.1                                                                                & 6.0 (1.5)                                                                                           & 17.5 (7.3)                                                                                          & 24.4 (8.1)                                                                                          & 11.7 (8.6)                                                                                          & map                        & $2.81^{+0.85}_{-0.69}$        & $12.67^{+0.26}_{-0.17}$                                & $2.84^{+0.26}_{-0.17}$                            \\
G12.13                     & 176.58833                                                                 & -0.16333                                                                   & 4.0                                                                                & 7.5 (1.9)                                                                                           & 0.5 (7.3)                                                                                           & 1.6 (8.1)                                                                                           & -10.2 (8.6)                                                                                         & map                        & $11.91^{+2.27}_{-3.36}$       & $13.18^{+0.25}_{-0.29}$                                & $3.35^{+0.25}_{-0.29}$                            \\
G12.14                     & 176.66611                                                                 & -0.23444                                                                   & 4.0                                                                                & 5.5 (1.4)                                                                                           & -7.3 (7.3)                                                                                          & -3.5 (8.1)                                                                                          & 1.1 (8.6)                                                                                           & map                        & $11.31^{+2.61}_{-3.51}$       & $13.04^{+0.23}_{-0.28}$                                & $3.21^{+0.23}_{-0.28}$                            \\
G12.15                     & 176.69611                                                                 & -0.11778                                                                   & 3.8                                                                                & 6.7 (1.8)                                                                                           & 31.0 (6.5)                                                                                          & 31.7 (7.5)                                                                                          & 27.2 (7.8)                                                                                          & cat                        & $2.39^{+0.72}_{-0.64}$        & $12.74^{+0.3}_{-0.17}$                                 & $2.92^{+0.3}_{-0.17}$                             \\ \hline
\multicolumn{12}{|c|}{Lockman}                                                                                                                                                                                                                                                                                                                                                                                                                                                                                                                                                                                                                                                                                                                                                                                                                                             \\ \hline
Lockman.01                 & 158.26016                                                                 & 59.15165                                                                   & 4.2                                                                                & 7.1 (1.7)                                                                                           & 21.1 (3.8)                                                                                          & 18.7 (4.8)                                                                                          & 9.4 (4.6)                                                                                           & cat                        & $2.52^{+0.69}_{-0.57}$        & $12.61^{+0.25}_{-0.2}$                                 & $2.78^{+0.25}_{-0.2}$                             \\
Lockman.02                 & 158.48582                                                                 & 59.23719                                                                   & 4.0                                                                                & 6.7 (1.7)                                                                                           & 20.0 (3.8)                                                                                          & 23.4 (4.8)                                                                                          & 26.3 (4.7)                                                                                          & cat                        & $3.05^{+0.73}_{-0.48}$        & $12.84^{+0.23}_{-0.17}$                                & $3.02^{+0.23}_{-0.17}$                            \\ \hline
\multicolumn{12}{|c|}{NGP1}                                                                                                                                                                                                                                                                                                                                                                                                                                                                                                                                                                                                                                                                                                                                                                                                                                                \\ \hline
NGP1.01                    & 201.11372                                                                 & 28.74656                                                                   & 33.8                                                                               & 42.8 (2.5)                                                                                          & 342.3 (5.6)                                                                                         & 371.0 (5.9)                                                                                         & 250.9 (6.9)                                                                                         & cat                        & $2.07^{+0.8}_{-1.54}$         & $13.62^{+0.42}_{-0.47}$                                & $3.79^{+0.42}_{-0.47}$                            \\
NGP1.02                    & 201.06683                                                                 & 28.74321                                                                   & 5.3                                                                                & 8.5 (1.7)                                                                                           & 3.7 (5.6)                                                                                           & 4.2 (5.9)                                                                                           & 3.5 (7.1)                                                                                           & map                        & $10.01^{+3.47}_{-3.44}$       & $13.17^{+0.26}_{-0.3}$                                 & $3.34^{+0.26}_{-0.3}$                            
\end{tabular}
\end{table}

\end{landscape}

\begin{landscape}

\begin{table}
\centering
\scriptsize
\contcaption{}
\label{tab:continued}
\begin{tabular}{cccccccccccc}
\hline
\multicolumn{1}{|c|}{Name} & \multicolumn{1}{c|}{\begin{tabular}[c]{@{}c@{}}RA\\ (J2000)\end{tabular}} & \multicolumn{1}{c|}{\begin{tabular}[c]{@{}c@{}}Dec\\ (J2000)\end{tabular}} & \multicolumn{1}{c|}{\begin{tabular}[c]{@{}c@{}}S/N \\ (at 850$\mu$m)\end{tabular}} & \multicolumn{1}{c|}{\begin{tabular}[c]{@{}c@{}}$S_{850}$ ($\sigma_{850}$)\\ {[}mJy{]}\end{tabular}} & \multicolumn{1}{c|}{\begin{tabular}[c]{@{}c@{}}$S_{250}$ ($\sigma_{250}$)\\ {[}mJy{]}\end{tabular}} & \multicolumn{1}{c|}{\begin{tabular}[c]{@{}c@{}}$S_{350}$ ($\sigma_{350}$)\\ {[}mJy{]}\end{tabular}} & \multicolumn{1}{c|}{\begin{tabular}[c]{@{}c@{}}$S_{500}$ ($\sigma_{500}$)\\ {[}mJy{]}\end{tabular}} & \multicolumn{1}{c|}{Class} & \multicolumn{1}{c|}{Redshift} & \multicolumn{1}{c|}{$log(L_{\mathrm{IR}}[L_{\odot}])$} & \multicolumn{1}{c|}{$log(SFR[M_{\odot}yr^{-1}])$} \\ \hline
NGP1.03                    & 201.09470                                                                 & 28.77211                                                                   & 5.1                                                                                & 7.0 (1.4)                                                                                           & 10.0 (5.6)                                                                                          & 4.3 (5.9)                                                                                           & 9.7 (7.1)                                                                                           & map                        & $6.83^{+4.06}_{-2.3}$         & $12.95^{+0.22}_{-0.27}$                                & $3.12^{+0.22}_{-0.27}$                            \\
NGP1.04                    & 201.12385                                                                 & 28.73989                                                                   & 4.9                                                                                & 5.1 (1.1)                                                                                           & 40.9 (5.8)                                                                                          & 25.9 (6.0)                                                                                          & 0.5 (7.1)                                                                                           & cat                        & $1.53^{+0.71}_{-0.39}$        & $12.47^{+0.38}_{-0.15}$                                & $2.64^{+0.38}_{-0.15}$                            \\
NGP1.05                    & 201.08455                                                                 & 28.78989                                                                   & 4.4                                                                                & 6.1 (1.4)                                                                                           & 74.3 (5.7)                                                                                          & 53.7 (5.8)                                                                                          & 26.0 (7.0)                                                                                          & cat                        & $1.4^{+0.5}_{-0.79}$          & $12.67^{+0.39}_{-0.35}$                                & $2.84^{+0.39}_{-0.35}$                            \\
NGP1.06                    & 201.07569                                                                 & 28.76544                                                                   & 4.3                                                                                & 6.0 (1.4)                                                                                           & -6.8 (5.6)                                                                                          & 1.0 (5.9)                                                                                           & 4.8 (7.1)                                                                                           & map                        & $10.55^{+3.04}_{-3.19}$       & $13.05^{+0.24}_{-0.25}$                                & $3.22^{+0.24}_{-0.25}$                            \\
NGP1.08                    & 201.05287                                                                 & 28.76543                                                                   & 4.1                                                                                & 6.0 (1.5)                                                                                           & 21.1 (5.6)                                                                                          & 17.8 (5.9)                                                                                          & 14.8 (7.1)                                                                                          & cat*                       & $2.75^{+0.85}_{-0.63}$        & $12.66^{+0.24}_{-0.19}$                                & $2.84^{+0.24}_{-0.19}$                            \\ \hline
\multicolumn{12}{|c|}{NGP2}                                                                                                                                                                                                                                                                                                                                                                                                                                                                                                                                                                                                                                                                                                                                                                                                                                                \\ \hline
NGP2.01                    & 199.87603                                                                 & 26.49655                                                                   & 5.9                                                                                & 12.5 (2.2)                                                                                          & 69.8 (4.3)                                                                                          & 72.9 (4.7)                                                                                          & 58.7 (5.9)                                                                                          & cat                        & $2.28^{+0.87}_{-0.64}$        & $13.09^{+0.34}_{-0.15}$                                & $3.26^{+0.34}_{-0.15}$                            \\
NGP2.02                    & 199.88225                                                                 & 26.41544                                                                   & 4.6                                                                                & 9.5 (2.1)                                                                                           & 25.9 (4.4)                                                                                          & 38.8 (4.8)                                                                                          & 33.6 (5.9)                                                                                          & cat                        & $3.05^{+0.72}_{-0.45}$        & $13.01^{+0.21}_{-0.21}$                                & $3.18^{+0.21}_{-0.21}$                            \\
NGP2.03                    & 199.90210                                                                 & 26.47433                                                                   & 4.2                                                                                & 6.3 (1.5)                                                                                           & 16.0 (4.4)                                                                                          & 17.5 (4.8)                                                                                          & 12.6 (6.0)                                                                                          & cat*                       & $3.05^{+0.82}_{-0.6}$         & $12.69^{+0.23}_{-0.2}$                                 & $2.86^{+0.23}_{-0.2}$                             \\
NGP2.04                    & 199.87232                                                                 & 26.42766                                                                   & 4.0                                                                                & 7.2 (1.8)                                                                                           & 53.5 (4.4)                                                                                          & 54.9 (4.7)                                                                                          & 50.4 (6.2)                                                                                          & cat*                       & $2.12^{+0.79}_{-0.81}$        & $12.86^{+0.38}_{-0.16}$                                & $3.03^{+0.38}_{-0.16}$                            \\
NGP2.05                    & 199.88969                                                                 & 26.44433                                                                   & 4.0                                                                                & 6.7 (1.7)                                                                                           & 7.3 (5.6)                                                                                           & 19.0 (5.9)                                                                                          & 23.1 (7.1)                                                                                          & map                        & $3.83^{+0.93}_{-0.75}$        & $12.8^{+0.2}_{-0.19}$                                  & $2.97^{+0.2}_{-0.19}$                             \\ \hline
\multicolumn{12}{|c|}{NGP3}                                                                                                                                                                                                                                                                                                                                                                                                                                                                                                                                                                                                                                                                                                                                                                                                                                                \\ \hline
NGP3.01($\#$)                    & 202.91419                                                                 & 23.69914                                                                   & 4.2                                                                                & 16.3 (4.0)                                                                                          & -3.0 (5.6)                                                                                          & 3.3 (5.9)                                                                                           & 6.0 (7.1)                                                                                           & map                        & $11.48^{+2.38}_{-2.94}$       & $13.51^{+0.23}_{-0.27}$                                & $3.69^{+0.23}_{-0.27}$                            \\
NGP3.02                    & 202.96153                                                                 & 23.74802                                                                   & 4.1                                                                                & 8.5 (2.1)                                                                                           & -3.2 (5.6)                                                                                          & 1.8 (5.9)                                                                                           & 9.6 (7.1)                                                                                           & map                        & $9.96^{+3.25}_{-3.13}$        & $13.16^{+0.25}_{-0.29}$                                & $3.33^{+0.25}_{-0.29}$                            \\
NGP3.03($\#$)                    & 202.86200                                                                 & 23.71912                                                                   & 4.1                                                                                & 11.1 (2.8)                                                                                          & 51.2 (5.6)                                                                                          & 63.5 (5.8)                                                                                          & 57.5 (7.2)                                                                                          & cat                        & $2.56^{+0.78}_{-0.38}$        & $13.1^{+0.28}_{-0.15}$                                 & $3.28^{+0.28}_{-0.15}$                            \\ \hline
\multicolumn{12}{|c|}{NGP4}                                                                                                                                                                                                                                                                                                                                                                                                                                                                                                                                                                                                                                                                                                                                                                                                                                                \\ \hline
NGP4.01                    & 198.55701                                                                 & 26.49932                                                                   & 5.3                                                                                & 11.7 (2.3)                                                                                          & 26.7 (5.8)                                                                                          & 38.3 (5.8)                                                                                          & 31.8 (7.5)                                                                                          & cat                        & $3.16^{+0.76}_{-0.42}$        & $13.04^{+0.21}_{-0.17}$                                & $3.21^{+0.21}_{-0.17}$                            \\
NGP4.02                    & 198.59425                                                                 & 26.55044                                                                   & 5.2                                                                                & 10.9 (2.1)                                                                                          & 47.8 (5.8)                                                                                          & 51.1 (5.9)                                                                                          & 44.2 (7.1)                                                                                          & cat                        & $2.41^{+0.76}_{-0.61}$        & $12.97^{+0.31}_{-0.15}$                                & $3.14^{+0.31}_{-0.15}$                            \\
NGP4.03                    & 198.56570                                                                 & 26.50822                                                                   & 4.8                                                                                & 8.3 (1.8)                                                                                           & 17.0 (5.6)                                                                                          & 21.2 (5.9)                                                                                          & 28.9 (7.1)                                                                                          & map                        & $5.65^{+4.27}_{-1.82}$        & $12.91^{+0.28}_{-0.24}$                                & $3.08^{+0.28}_{-0.24}$                            \\
NGP4.04                    & 198.57936                                                                 & 26.49155                                                                   & 4.5                                                                                & 7.2 (1.6)                                                                                           & 15.4 (5.5)                                                                                          & 7.4 (6.0)                                                                                           & 4.5 (7.0)                                                                                           & cat                        & $4.63^{+3.5}_{-1.34}$         & $12.81^{+0.24}_{-0.23}$                                & $2.99^{+0.24}_{-0.23}$                            \\
NGP4.05                    & 198.59798                                                                 & 26.50267                                                                   & 4.4                                                                                & 9.7 (2.3)                                                                                           & 35.0 (5.8)                                                                                          & 31.4 (5.8)                                                                                          & 30.1 (6.9)                                                                                          & cat                        & $12.47^{+1.97}_{-2.33}$       & $13.32^{+0.22}_{-0.28}$                                & $3.49^{+0.22}_{-0.28}$                            \\
NGP4.06                    & 198.55701                                                                 & 26.51710                                                                   & 4.2                                                                                & 8.0 (1.9)                                                                                           & -4.5 (5.6)                                                                                          & 7.6 (5.9)                                                                                           & -0.4 (7.1)                                                                                          & map                        & $3.36^{+0.82}_{-0.51}$        & $12.88^{+0.21}_{-0.16}$                                & $3.06^{+0.21}_{-0.16}$                            \\
NGP4.07                    & 198.67126                                                                 & 26.54599                                                                   & 4.1                                                                                & 6.7 (1.7)                                                                                           & -10.7 (5.6)                                                                                         & -14.6 (5.9)                                                                                         & -13.6 (7.1)                                                                                         & map                        & $10.9^{+2.99}_{-3.28}$        & $13.08^{+0.27}_{-0.28}$                                & $3.25^{+0.27}_{-0.28}$                            \\
NGP4.08                    & 198.60419                                                                 & 26.52156                                                                   & 3.9                                                                                & 8.1 (2.1)                                                                                           & 5.2 (5.6)                                                                                           & 25.3 (5.9)                                                                                          & 30.8 (7.1)                                                                                          & map                        & $9.88^{+3.25}_{-3.05}$        & $13.12^{+0.24}_{-0.26}$                                & $3.29^{+0.24}_{-0.26}$                            \\
NGP4.09                    & 198.54955                                                                 & 26.52377                                                                   & 3.8                                                                                & 7.2 (1.9)                                                                                           & -1.2 (5.6)                                                                                          & -6.9 (5.9)                                                                                          & -16.9 (7.1)                                                                                         & map                        & $12.76^{+1.54}_{-2.9}$        & $13.16^{+0.23}_{-0.3}$                                 & $3.33^{+0.23}_{-0.3}$                             \\ \hline
\multicolumn{12}{|c|}{NGP5}                                                                                                                                                                                                                                                                                                                                                                                                                                                                                                                                                                                                                                                                                                                                                                                                                                                \\ \hline
NGP5.01                    & 205.17118                                                                 & 32.61808                                                                   & 8.2                                                                                & 13.8 (1.8)                                                                                          & 30.6 (4.8)                                                                                          & 53.5 (5.5)                                                                                          & 50.0 (6.5)                                                                                          & cat*                       & $3.19^{+0.7}_{-0.34}$         & $13.16^{+0.22}_{-0.17}$                                & $3.33^{+0.22}_{-0.17}$                            \\
NGP5.02                    & 205.18570                                                                 & 32.64919                                                                   & 5.9                                                                                & 10.3 (1.8)                                                                                          & 34.0 (4.9)                                                                                          & 32.8 (5.2)                                                                                          & 30.6 (6.3)                                                                                          & cat                        & $2.69^{+0.68}_{-0.41}$        & $12.91^{+0.25}_{-0.17}$                                & $3.09^{+0.25}_{-0.17}$                            \\
NGP5.03                    & 205.17646                                                                 & 32.59808                                                                   & 5.2                                                                                & 9.1 (1.8)                                                                                           & 0.8 (5.6)                                                                                           & 13.0 (5.9)                                                                                          & 37.7 (7.1)                                                                                          & map                        & $4.79^{+1.17}_{-0.91}$        & $13.0^{+0.17}_{-0.19}$                                 & $3.17^{+0.17}_{-0.19}$                            \\
NGP5.04                    & 205.10917                                                                 & 32.63362                                                                   & 4.4                                                                                & 9.3 (2.2)                                                                                           & 36.0 (4.9)                                                                                          & 38.1 (5.3)                                                                                          & 33.8 (6.4)                                                                                          & cat                        & $2.58^{+0.68}_{-0.48}$        & $12.92^{+0.25}_{-0.17}$                                & $3.09^{+0.25}_{-0.17}$                            \\
NGP5.05                    & 205.19624                                                                 & 32.61030                                                                   & 4.2                                                                                & 6.5 (1.6)                                                                                           & -2.9 (5.6)                                                                                          & -0.2 (5.9)                                                                                          & 9.9 (7.1)                                                                                           & map                        & $9.86^{+3.32}_{-3.28}$        & $13.03^{+0.25}_{-0.29}$                                & $3.21^{+0.25}_{-0.29}$                            \\
NGP5.06                    & 205.18174                                                                 & 32.66031                                                                   & 4.0                                                                                & 6.6 (1.7)                                                                                           & 30.7 (5.1)                                                                                          & 39.3 (5.2)                                                                                          & 36.0 (6.4)                                                                                          & cat                        & $2.46^{+0.89}_{-0.71}$        & $12.81^{+0.34}_{-0.15}$                                & $2.98^{+0.34}_{-0.15}$                            \\
NGP5.07                    & 205.20811                                                                 & 32.60363                                                                   & 3.8                                                                                & 6.9 (1.8)                                                                                           & 12.6 (4.8)                                                                                          & 3.3 (5.2)                                                                                           & 1.5 (6.4)                                                                                           & cat*                       & $7.49^{+4.7}_{-3.31}$         & $12.9^{+0.33}_{-0.32}$                                 & $3.08^{+0.33}_{-0.32}$                            \\ \hline
\multicolumn{12}{|c|}{NGP6}                                                                                                                                                                                                                                                                                                                                                                                                                                                                                                                                                                                                                                                                                                                                                                                                                                                \\ \hline
NGP6.01                    & 200.81323                                                                 & 33.38511                                                                   & 5.4                                                                                & 4.5 (0.9)                                                                                           & 5.6 (5.6)                                                                                           & -3.4 (5.9)                                                                                          & 10.0 (7.1)                                                                                          & map                        & $9.45^{+3.74}_{-4.18}$        & $12.87^{+0.26}_{-0.3}$                                 & $3.05^{+0.26}_{-0.3}$                             \\
NGP6.02                    & 200.74136                                                                 & 33.40399                                                                   & 5.0                                                                                & 10.1 (2.1)                                                                                          & -5.5 (5.6)                                                                                          & 3.6 (5.9)                                                                                           & 7.5 (7.1)                                                                                           & map                        & $10.35^{+3.0}_{-2.94}$        & $13.26^{+0.25}_{-0.27}$                                & $3.43^{+0.25}_{-0.27}$                            \\
NGP6.03                    & 200.80258                                                                 & 33.38511                                                                   & 4.4                                                                                & 5.3 (1.2)                                                                                           & 49.0 (5.6)                                                                                          & 45.5 (5.8)                                                                                          & 18.3 (7.2)                                                                                          & cat                        & $1.61^{+0.45}_{-0.97}$        & $12.62^{+0.35}_{-0.4}$                                 & $2.79^{+0.35}_{-0.4}$                             \\
NGP6.04                    & 200.83452                                                                 & 33.39400                                                                   & 4.2                                                                                & 4.7 (1.1)                                                                                           & 35.8 (5.5)                                                                                          & 23.3 (5.9)                                                                                          & 14.8 (7.2)                                                                                          & cat                        & $1.72^{+0.69}_{-0.53}$        & $12.49^{+0.35}_{-0.17}$                                & $2.67^{+0.35}_{-0.17}$                            \\
NGP6.05                    & 200.80791                                                                 & 33.40956                                                                   & 4.2                                                                                & 5.4 (1.3)                                                                                           & 30.1 (5.8)                                                                                          & 32.9 (6.0)                                                                                          & 17.0 (7.4)                                                                                          & cat                        & $2.1^{+0.65}_{-0.63}$         & $12.62^{+0.3}_{-0.16}$                                 & $2.79^{+0.3}_{-0.16}$                             \\
NGP6.06                    & 200.76264                                                                 & 33.43288                                                                   & 4.1                                                                                & 6.6 (1.6)                                                                                           & 92.5 (5.6)                                                                                          & 67.6 (5.8)                                                                                          & 40.2 (6.9)                                                                                          & cat                        & $1.4^{+0.55}_{-0.76}$         & $12.78^{+0.4}_{-0.36}$                                 & $2.96^{+0.4}_{-0.36}$                             \\
NGP6.09                    & 200.86513                                                                 & 33.40065                                                                   & 3.9                                                                                & 5.7 (1.5)                                                                                           & 3.8 (5.6)                                                                                           & 15.8 (5.9)                                                                                          & 32.1 (7.1)                                                                                          & map                        & $3.89^{+0.96}_{-0.71}$        & $12.78^{+0.19}_{-0.17}$                                & $2.95^{+0.19}_{-0.17}$                            \\
NGP6.12                    & 200.76397                                                                 & 33.42399                                                                   & 3.8                                                                                & 6.9 (1.9)                                                                                           & 33.0 (5.4)                                                                                          & 20.7 (6.0)                                                                                          & 14.8 (7.1)                                                                                          & cat                        & $2.12^{+0.64}_{-0.64}$        & $12.62^{+0.27}_{-0.23}$                                & $2.8^{+0.27}_{-0.23}$                            
\end{tabular}
\end{table}

\end{landscape}

\begin{landscape}

\begin{table}
\centering
\scriptsize
\contcaption{}
\label{tab:continued2}
\begin{tabular}{cccccccccccc}
\hline
\multicolumn{1}{|c|}{Name} & \multicolumn{1}{c|}{\begin{tabular}[c]{@{}c@{}}RA\\ (J2000)\end{tabular}} & \multicolumn{1}{c|}{\begin{tabular}[c]{@{}c@{}}Dec\\ (J2000)\end{tabular}} & \multicolumn{1}{c|}{\begin{tabular}[c]{@{}c@{}}S/N \\ (at 850$\mu$m)\end{tabular}} & \multicolumn{1}{c|}{\begin{tabular}[c]{@{}c@{}}$S_{850}$ ($\sigma_{850}$)\\ {[}mJy{]}\end{tabular}} & \multicolumn{1}{c|}{\begin{tabular}[c]{@{}c@{}}$S_{250}$ ($\sigma_{250}$)\\ {[}mJy{]}\end{tabular}} & \multicolumn{1}{c|}{\begin{tabular}[c]{@{}c@{}}$S_{350}$ ($\sigma_{350}$)\\ {[}mJy{]}\end{tabular}} & \multicolumn{1}{c|}{\begin{tabular}[c]{@{}c@{}}$S_{500}$ ($\sigma_{500}$)\\ {[}mJy{]}\end{tabular}} & \multicolumn{1}{c|}{Class} & \multicolumn{1}{c|}{Redshift} & \multicolumn{1}{c|}{$log(L_{\mathrm{IR}}[L_{\odot}])$} & \multicolumn{1}{c|}{$log(SFR[M_{\odot}yr^{-1}])$} \\ \hline
\multicolumn{12}{|c|}{NGP7}                                                                                                                                                                                                                                                                                                                                                                                                                                                                                                                                                                                                                                                                                                                                                                                                                                                \\ \hline
NGP7.01                    & 204.25680                                                                 & 32.14564                                                                   & 6.6                                                                                & 11.6 (1.8)                                                                                          & 59.2 (5.6)                                                                                          & 74.5 (5.7)                                                                                          & 67.1 (7.3)                                                                                          & cat                        & $2.35^{+0.98}_{-0.8}$         & $13.09^{+0.34}_{-0.19}$                                & $3.26^{+0.34}_{-0.19}$                            \\
NGP7.02                    & 204.23449                                                                 & 32.13785                                                                   & 4.8                                                                                & 11.4 (2.4)                                                                                          & 30.3 (5.8)                                                                                          & 29.9 (5.8)                                                                                          & 16.6 (7.4)                                                                                          & cat                        & $2.79^{+0.79}_{-0.54}$        & $12.9^{+0.21}_{-0.23}$                                 & $3.07^{+0.21}_{-0.23}$                            \\
NGP7.03                    & 204.29617                                                                 & 32.18119                                                                   & 4.7                                                                                & 9.2 (2.0)                                                                                           & 24.1 (5.7)                                                                                          & 26.5 (5.8)                                                                                          & 20.3 (7.6)                                                                                          & cat                        & $2.97^{+0.77}_{-0.54}$        & $12.86^{+0.22}_{-0.2}$                                 & $3.03^{+0.22}_{-0.2}$                             \\
NGP7.04                    & 204.27517                                                                 & 32.14119                                                                   & 4.6                                                                                & 6.6 (1.5)                                                                                           & 23.6 (5.7)                                                                                          & 27.7 (5.8)                                                                                          & 17.7 (7.4)                                                                                          & cat                        & $2.56^{+0.85}_{-0.54}$        & $12.71^{+0.3}_{-0.15}$                                 & $2.88^{+0.3}_{-0.15}$                             \\
NGP7.05                    & 204.34604                                                                 & 32.15229                                                                   & 4.6                                                                                & 10.8 (2.4)                                                                                          & 19.1 (5.6)                                                                                          & 32.0 (5.9)                                                                                          & 29.8 (7.1)                                                                                          & map                        & $3.47^{+0.77}_{-0.51}$        & $13.01^{+0.19}_{-0.19}$                                & $3.19^{+0.19}_{-0.19}$                            \\
NGP7.06                    & 204.31060                                                                 & 32.13675                                                                   & 4.6                                                                                & 7.6 (1.7)                                                                                           & 4.8 (5.6)                                                                                           & 12.2 (5.9)                                                                                          & 11.8 (7.1)                                                                                          & map                        & $5.59^{+2.88}_{-1.54}$        & $12.9^{+0.23}_{-0.22}$                                 & $3.07^{+0.23}_{-0.22}$                            \\
NGP7.07                    & 204.23058                                                                 & 32.08896                                                                   & 4.3                                                                                & 8.1 (2.0)                                                                                           & 6.4 (5.6)                                                                                           & 1.6 (5.9)                                                                                           & -13.1 (7.1)                                                                                         & map                        & $11.85^{+2.27}_{-3.18}$       & $13.22^{+0.21}_{-0.3}$                                 & $3.4^{+0.21}_{-0.3}$                              \\
NGP7.08                    & 204.29222                                                                 & 32.07564                                                                   & 4.2                                                                                & 13.1 (3.2)                                                                                          & 17.8 (5.6)                                                                                          & 23.1 (6.0)                                                                                          & 10.1 (7.1)                                                                                          & cat*                       & $3.8^{+0.99}_{-0.93}$         & $12.96^{+0.21}_{-0.27}$                                & $3.13^{+0.21}_{-0.27}$                            \\
NGP7.09                    & 204.28173                                                                 & 32.16675                                                                   & 4.1                                                                                & 6.4 (1.6)                                                                                           & 2.7 (5.6)                                                                                           & 12.7 (5.9)                                                                                          & 28.0 (7.1)                                                                                          & map                        & $4.35^{+1.35}_{-0.89}$        & $12.82^{+0.2}_{-0.19}$                                 & $2.99^{+0.2}_{-0.19}$                             \\
NGP7.10                    & 204.23973                                                                 & 32.16008                                                                   & 4.0                                                                                & 6.7 (1.7)                                                                                           & 36.6 (5.7)                                                                                          & 31.4 (5.9)                                                                                          & 19.9 (7.1)                                                                                          & cat                        & $2.08^{+0.56}_{-0.59}$        & $12.67^{+0.3}_{-0.15}$                                 & $2.85^{+0.3}_{-0.15}$                             \\
NGP7.11                    & 204.29485                                                                 & 32.15342                                                                   & 4.0                                                                                & 5.7 (1.5)                                                                                           & 46.6 (5.8)                                                                                          & 55.7 (5.9)                                                                                          & 55.7 (7.6)                                                                                          & cat                        & $2.01^{+0.67}_{-1.41}$        & $12.81^{+0.33}_{-0.57}$                                & $2.98^{+0.33}_{-0.57}$                            \\
NGP7.12                    & 204.30403                                                                 & 32.11897                                                                   & 3.9                                                                                & 6.0 (1.6)                                                                                           & 1.7 (5.6)                                                                                           & 0.8 (5.9)                                                                                           & 3.3 (7.1)                                                                                           & map                        & $10.3^{+3.31}_{-3.74}$        & $13.0^{+0.28}_{-0.3}$                                  & $3.17^{+0.28}_{-0.3}$                             \\
NGP7.13                    & 204.24760                                                                 & 32.18563                                                                   & 3.9                                                                                & 7.7 (2.0)                                                                                           & -11.0 (5.6)                                                                                         & -11.3 (5.9)                                                                                         & -17.0 (7.1)                                                                                         & map                        & $12.74^{+1.64}_{-2.75}$       & $13.21^{+0.22}_{-0.31}$                                & $3.38^{+0.22}_{-0.31}$                            \\ \hline
\multicolumn{12}{|c|}{NGP8}                                                                                                                                                                                                                                                                                                                                                                                                                                                                                                                                                                                                                                                                                                                                                                                                                                                \\ \hline
NGP8.01                    & 202.31798                                                                 & 28.21872                                                                   & 5.0                                                                                & 10.2 (2.1)                                                                                          & 29.2 (5.7)                                                                                          & 14.0 (6.2)                                                                                          & -0.9 (7.4)                                                                                          & cat*                       & $2.53^{+0.88}_{-0.71}$        & $12.69^{+0.22}_{-0.27}$                                & $2.86^{+0.22}_{-0.27}$                            \\
NGP8.02                    & 202.36968                                                                 & 28.23650                                                                   & 4.5                                                                                & 7.7 (1.8)                                                                                           & 49.6 (5.7)                                                                                          & 64.7 (6.0)                                                                                          & 65.6 (6.9)                                                                                          & cat                        & $2.27^{+0.97}_{-0.96}$        & $12.97^{+0.36}_{-0.25}$                                & $3.15^{+0.36}_{-0.25}$                            \\
NGP8.03                    & 202.39237                                                                 & 28.18538                                                                   & 4.1                                                                                & 9.1 (2.3)                                                                                           & 36.4 (5.6)                                                                                          & 38.3 (6.0)                                                                                          & 35.8 (7.1)                                                                                          & cat                        & $2.56^{+0.77}_{-0.47}$        & $12.91^{+0.27}_{-0.17}$                                & $3.09^{+0.27}_{-0.17}$                            \\
NGP8.04                    & 202.33564                                                                 & 28.19761                                                                   & 3.8                                                                                & 7.6 (2.0)                                                                                           & 5.4 (5.6)                                                                                           & 10.7 (5.9)                                                                                          & -9.4 (7.1)                                                                                          & map                        & $10.49^{+3.24}_{-4.02}$       & $13.08^{+0.28}_{-0.33}$                                & $3.26^{+0.28}_{-0.33}$                            \\
NGP8.05                    & 202.29527                                                                 & 28.24426                                                                   & 3.7                                                                                & 8.2 (2.3)                                                                                           & 40.6 (5.7)                                                                                          & 41.8 (6.0)                                                                                          & 25.4 (6.9)                                                                                          & cat                        & $2.21^{+0.7}_{-0.55}$         & $12.81^{+0.3}_{-0.16}$                                 & $2.99^{+0.3}_{-0.16}$                             \\ \hline
\multicolumn{12}{|c|}{NGP9}                                                                                                                                                                                                                                                                                                                                                                                                                                                                                                                                                                                                                                                                                                                                                                                                                                                \\ \hline
NGP9.01                    & 194.74984                                                                 & 31.58490                                                                   & 3.9                                                                                & 8.1 (2.1)                                                                                           & 5.1 (5.6)                                                                                           & 10.9 (5.9)                                                                                          & 13.9 (7.1)                                                                                          & map                        & $5.7^{+3.18}_{-1.52}$         & $12.93^{+0.24}_{-0.24}$                                & $3.11^{+0.24}_{-0.24}$                            \\
NGP9.02                    & 194.87636                                                                 & 31.59490                                                                   & 3.9                                                                                & 7.3 (1.9)                                                                                           & 27.7 (5.6)                                                                                          & 25.4 (6.2)                                                                                          & 26.8 (7.7)                                                                                          & cat                        & $2.63^{+0.7}_{-0.57}$         & $12.78^{+0.24}_{-0.17}$                                & $2.95^{+0.24}_{-0.17}$                            \\
NGP9.03                    & 194.86857                                                                 & 31.64379                                                                   & 3.9                                                                                & 6.0 (1.6)                                                                                           & 18.9 (5.5)                                                                                          & 6.2 (5.7)                                                                                           & 9.4 (6.8)                                                                                           & cat                        & $3.3^{+1.59}_{-1.07}$         & $12.63^{+0.23}_{-0.25}$                                & $2.8^{+0.23}_{-0.25}$                             \\
NGP9.04                    & 194.82157                                                                 & 31.53603                                                                   & 3.8                                                                                & 8.2 (2.2)                                                                                           & 17.4 (5.8)                                                                                          & -5.5 (6.0)                                                                                          & -7.6 (7.7)                                                                                          & cat*                       & $11.42^{+2.59}_{-4.49}$       & $13.19^{+0.26}_{-0.39}$                                & $3.36^{+0.26}_{-0.39}$                           
\end{tabular}
\end{table}

\end{landscape}


\bsp	
\label{lastpage}
\end{document}